\documentclass[conference]{IEEEtran}
\IEEEoverridecommandlockouts

\usepackage{tikz}
\usepackage{amsmath}
\usepackage{listings}
\usepackage{adjustbox}
\usepackage{longtable}
\definecolor{darkgreen}{rgb}{0.0, 0.42, 0.24}
\definecolor{LightGray}{gray}{0.97}
\definecolor{my}{rgb}{0.59, 0.78, 0.64}
\colorlet{mine}{my!15}
\definecolor{earthyellow}{rgb}{0.88, 0.66, 0.37}
\colorlet{mine2}{earthyellow!15}
\definecolor{deepcarrotorange}{rgb}{0.91, 0.41, 0.17}
\colorlet{mine3}{deepcarrotorange!15}
\usepackage{colortbl}
\usepackage{booktabs}
\usepackage{multirow}
\usepackage{amssymb}
\usepackage{pifont}
\newcommand{\cmark}{\ding{51}}%
\newcommand{\xmark}{\ding{55}}%
\usepackage{subcaption}
\usepackage{hyperref}
\usepackage{soul}
\usepackage[para,online,flushleft]{threeparttable}
\def\BibTeX{{\rm B\kern-.05em{\sc i\kern-.025em b}\kern-.08em
    T\kern-.1667em\lower.7ex\hbox{E}\kern-.125emX}}
\begin{document}
\date{}
\title{Do you \emph{really} code? Designing and Evaluating Screening Questions for\\ Online Surveys with Programmers}
\author{\IEEEauthorblockN{Anastasia Danilova}
\IEEEauthorblockA{
\textit{University of Bonn}\\
Bonn, Germany \\
danilova@cs.uni-bonn.de}
\and
\IEEEauthorblockN{Alena Naiakshina}
\IEEEauthorblockA{
\textit{University of Bonn}\\
Bonn, Germany  \\
naiakshi@cs.uni-bonn.de}
\and
\IEEEauthorblockN{Stefan Horstmann}
\IEEEauthorblockA{
\textit{University of Bonn}\\
Bonn, Germany \\
Stefan.Horstmann@gmx.net}
\and
\IEEEauthorblockN{Matthew Smith}
\IEEEauthorblockA{
\textit{University of Bonn, Fraunhofer FKIE}\\
Bonn, Germany  \\
smith@cs.uni-bonn.de}

}



\maketitle

\begin{abstract}
Recruiting professional programmers in sufficient numbers for research studies can be challenging because they often cannot spare the time, or due to their geographical distribution and potentially the cost involved. Online platforms such as Clickworker or Qualtrics do provide options to recruit participants with programming skill; however, misunderstandings and fraud can be an issue. 
This can result in participants without programming skill taking part in studies and surveys. If these participants are not detected, they can cause detrimental noise in the survey data. In this paper, we develop screener questions that are easy and quick to answer for people with programming skill but difficult to answer correctly for those without. In order to evaluate our questionnaire for efficacy and efficiency, we recruited several batches of participants with and without programming skill and tested the questions. In our batch 42\% of Clickworkers stating that they have programming skill did not meet our criteria and we would recommend filtering these from studies. 
We also evaluated the questions in an adversarial setting. We conclude with a set of recommended questions which researchers can use to recruit participants with programming skill from online platforms.

\end{abstract}

\section{Introduction}
Conducting user studies is an essential part of empirical software engineering; however, recruiting enough participants with programming skill can be challenging. Researchers use different methods to recruit programmers for studies related to software engineering. One common method is to recruit computer science (CS) students or developers from local companies for a local and in-person study. On the other end of the spectrum, researchers also commonly recruit developers on an international level for remote studies, using a variety of platforms. One advantage of recruiting locally is that researchers can be fairly certain that most of their sample actually has a computer science/programming background (e.g., \cite{naiakshina2017developers, naiakshina2018deception, acar2016you, ponzanelli2017supporting, williams2019mercury, chen2016towards}). Unfortunately, finding enough willing participants locally can be difficult; Naiakshina et al.~\cite{naiakshina2018deception} reported that they could only get 40 out of 1600 CS students to take part in a study where the compensation was 100 euros. To widen the recruitment pool and include non-student participants, it is common for researchers to resort to online studies and recruit participants online (e.g., \cite{wijayarathna2018johnny, sousa2017software, gorski2018developers, acar2017security, naiakshina2019if, bau2012vulnerability, Assal:2019:TSB:3290605.3300519, yamashita2013developers, naiakshina2020conducting}). Diverse recruitment strategies have been used, such as cold-calling programmers on platforms such as Stack Overflow, GitHub, Meet-up groups, etc. or posting open invitations on social media, in forums, newsletters and events, with the expectation being that participants without programming knowledge will not sign up for the studies~\cite{Assal:2019:TSB:3290605.3300519, beller2018dichotomy, spadini2020primers}.
However, since researchers often offer significantly higher compensation than for end-user studies~\cite{naiakshina2017developers,naiakshina2018deception,naiakshina2019if}, there can be an incentive for participants to take part in a study despite having no programming skill. 

The acquisition of skill is divided in three overlapping phases: (1) knowledge acquisition, (2) knowledge association, and (3) autonomous task performance~\cite{fitts1967human, bergersen2014construction}. For \emph{programming skill}, we use the definition of Bergersen et al.~\cite{bergersen2014construction}, which is in accordance to the definition used in psychology~\cite{fitts1967human, shute1991whois, anderson1987skill, anderson1989skill, anderson1984learning}: ``the ability to use one’s knowledge effectively and readily in execution or performance of programming tasks.'' In previous work with programmers, participants were often expected to have programming skill in various programming languages such as Python, Java, C, Perl, Haskell, JavaScript, PHP, etc. (e.g.,~\cite{acar2016you, wijayarathna2018johnny, gorski2018developers, acar2017security, naiakshina2020conducting, bergersen2014construction, acar2017comparing, balebako2014privacy, prechelt2010platforms, nadi2016jumping}). However, people who use programs such as Excel, manage a content management system (CMS) or write HTML might consider themselves as having programming skill. While this does not match our understanding and the requirements of previous research with programmers, we want to make it clear that participants stating that they have programming skill might be an honest misunderstanding from our perspective and not necessarily malicious intent. None the less, having these kinds of participants in programming related studies and surveys is detrimental. 


In studies which contain actual programming tasks, these participants can be detected fairly easily. However, since it is common practice to pay participants independently of how well they perform, they still cost time and cause financial damage to the researchers. More critically, studies which aim to examine attitudes towards software engineering related topics, such as new features of programming languages, API design, error handling, or security and privacy, might corrupt their data by including answers from participants who do not understand the subject matter because they do not have any programming skill. Examples for such studies are \cite{Assal:2019:TSB:3290605.3300519, yamashita2013developers, beller2018dichotomy, spadini2020primers, votipka2020building, danilova2020one, witschey2015quantifying, sheth2014us, nadi2016jumping, iacono2017and, oltrogge2015pin, oliveira2018api, balebako2014privacy}. 

Researchers can also use online recruitment platforms such as Clickworker~\cite{Clickworker} or Qualtrics~\cite{qualtrics} for developer studies (e.g.,~\cite{Assal:2019:TSB:3290605.3300519, danilova2020one}). These platforms provide panels to recruit participants with specific skills, such as programming. However, as mentioned above, there might be misunderstandings concerning the term and the compensation for participants with (self-stated) programming skills is higher than for other types of study participants, and consequently there is the risk of participants falsely stating that they have these skills. Danilova et al.~\cite{danilova2020one} came across such a case. They ran a survey to study developers' attitudes to security warning design, for which they used Qualtrics to recruit participants with programming skill. To test the participants' basic developer knowledge, the authors presented participants with some pseudocode that printed ``hello world'' backwards and asked what the output would be.
The authors reported that out of the 129 participants recruited on Qualtrics who stated that they had programming skill, only 33 gave the correct answer. Since most of the rest of the study consisted of closed questions, without the programming skill question, the authors would have included many participants in their data, who were not able to understand a very simple program. 

To enable researchers to be more confident about conducting online surveys with programmers, we investigated the research question:
\textit{Which questions can be used to screen out participants without any programming skill (while meeting our requirements of effectiveness, efficiency, and robustness against cheating)}?

We created a survey instrument consisting of questions which can be used to assess whether a participant actually has programming skill.
The questions are designed to take as little time as possible so they can be used in free or cheap pre-screening surveys. They needed to be so easy as to not annoy or challenge people with programming skill so we do  not falsely reject participants, but also hard enough so that people without programming skill cannot make an educated guess or google the answer quickly.

We designed 16 questions of different types and tested them with different groups of participants: CS students, professional programmers, students enrolled in a behavioral economics study platform, as well as participants from Clickworker with and without self-stated programming skill. We evaluated the results from these groups to provide a shortlist of questions which seemed promising as screening questions for developer studies. To evaluate these questions, we tested them in an adversarial environment, where we offered non-programmers an extra monetary incentive to answer them correctly by any means, including using the Internet. As a final result, we offer a list of questions we believe can be used to screen non-programmers out of surveys with only minimal overhead for the participants with actual programming skill.

\section{Related Work}
In order to provide context for this field of work, we separate our related work into two sections. We first outline developer studies, where recruitment strategies are discussed or chosen in a way that developers are targeted with a high probability. Second, we provide insight into other work that either uses identification and verification mechanisms in the research or investigates how participants could be tested for programming skill.


\subsection{Recruitment Strategies}
Researchers mostly designed the recruitment process in such a way that only software developers were targeted, such as through recruitment of participants on GitHub~\cite{wijayarathna2018johnny, gorski2018developers, acar2017security, graziotin2017unhappy, acar2017comparing} or on freelance platforms for developers~\cite{yamashita2013developers, yamashita2013surveying, bau2012vulnerability, naiakshina2019if}. 
For example, when investigating the effect of happiness on productivity, Graziotin et al.~\cite{graziotin2017unhappy}  recruited their participants by contacting software developers on GitHub~\cite{GitHub}. GitHub developers were also recruited for a security developer study conducted by Acar et al.~\cite{acar2017security} to investigate how they perform with regard to security-related tasks. Furthermore, Meyer et al.~\cite{meyer2019today}, when conducting a study on job satisfaction and productivity of software developers, directly recruited employees from Microsoft to ensure that all participants indeed work as software developers.

A number of recruitment strategies were compared by Baltes and Diehl in~\cite{baltes2016worse}. They recruited survey participants using different approaches: via a personal network, online networks and communities, directly contacting companies, public media, survey advertisement by software engineers, and by using information from GHTorrent~\cite{GHTorrent}, which collects data on public GitHub projects. 
They concluded that the most efficient way to recruit is to use public media and asking software engineers to advertise the survey. Nevertheless, Baltes and Diehl did raise the question as to whether recruitment strategies like commercial recruiting services or crowdsourcing platforms (e.g., Amazon Mechanical Turk) are suitable for developer recruitment, but did not follow up on it, which we do.
Another concern is also raised by Yamashita and Moonen~\cite{yamashita2013surveying}, who discussed the recruitment of freelance developers for surveys on the platform Freelancer.com. The authors acknowledged that employers rely on self-reported skills by members of such platforms. They suggested running skill assessment tests to ensure the internal validity of research findings. Our work addresses this issue. 

\subsection{Identifying Programming Skill}
Bergersen et al.~\cite{bergersen2014construction} constructed and validated an instrument for measuring Java programming skill. This was done by evaluating participants' performance on programming tasks. In a study lasting two days, 65 professional developers solved 19 Java programming tasks. Each task had a time limit of between 10 and 45 minutes. The research goal of Bergersen et al. was to construct an instrument to find the best programmers when recruiting for a job or allocating to projects. The task time of 10-45 minutes is too long for our purpose of screening participants before a survey. 


Feigenspan et al.~\cite{feigenspan2012measuring} conducted an experiment with 125 students and compared the self-reported programming experience with the participants' performance in solving program comprehension tasks. They found a correlation between self-reported experience and performance using a 40 minute survey instrument.
In addition to that, the authors highlighted that the software-engineering community is lacking a clear definition of \emph{programming experience}. However, by using this term, researchers often referred to years a participant was programming by using a specific programming language or in general. Interestingly, Bergersen et al.~\cite{bergersen2014construction} found a correlation between programming experience and skill, which was largest during the first year of experience and disappeared after about four years. 
While we also used program comprehension tasks in our study, we did not aim to understand how experience correlates with performance; our objective was to find a simple way to exclude participants without programming skill from online developer studies.

In~\cite{balebako2014privacy}, Balebako et al. conducted a study concerning the behavior of smartphone app developers in regard to security and privacy. 
Potential participants had to fill out a screening survey to qualify for the interviews. This survey included two technical questions to test for knowledge of app development. In addition, the authors verified their findings in a follow-up online survey with participants recruited on various online forums. Again, knowledge and attention check questions were required to be correctly answered by participants. Of the 460 responses, 232 had to be discarded because they did not fulfill the requirements. 
We contacted Balebako et al. and asked for their used screening questions. Unfortunately, the authors were not able to provide us the exact questions. However, they remembered to have asked for IDEs participants have experience with and apps they developed. After that the authors manually inspected whether the mentioned IDEs or apps exist. The authors acknowledged that especially the IDE question was hard to verify, because of the high amount of IDEs existing ``in the wild.'' Instead of asking for IDEs, we therefore decided to provide participants a list of programming languages including non-existing ones.

In~\cite{acar2016you}, Acar et al. studied the effect of information resources for software development on security by conducting an online survey and a lab study. For the online survey, the researchers sent about 50,000 emails to developers they were able to identify on Google Play. Of the 302 participants who completed the survey, Acar et al. excluded 7 for invalid answers. Additional filters to find non-developer participants were not applied. In the lab study, participants were recruited based on their experience with app development, having either completed a course on Android development or with work experience of at least one year. 
The participants first had to complete a short programming task to demonstrate their skills. However, following complaints that the task took too much time, they were instead tested with 5 multiple choice questions covering basic Android development knowledge, at least 3 of which needed to be answered correctly. This is a good example showing that programming tasks are not well suited as screening question and that multiple choice questions are more acceptable. 
We contacted Acar et al. and learned that their screening questions specifically covered Android development knowledge. Since we aimed to identify questions for general programming skill, we did not include them to our instrument. However, we included questions affecting developers beyond Android development, such as error handling.

Danilova et al.~\cite{danilova2020one} investigated the developers' preferences about security warnings in IDEs and tested participants' programming skill with a simple pseudocode multiple choice question. A detailed description of the study can be found in the Introduction. To further evaluate the used question, we included it to our question set as Q16.

Assal and Chiasson~\cite{Assal:2019:TSB:3290605.3300519} conducted online surveys to investigate developers' software security processes. A section of their participants were recruited through a paid service provided by Qualtrics. During survey completion, participants were provided different descriptions of software security and were prevented from progressing till they chose the authors' preferred definition of security. Participants initially providing incorrect answers were not excluded from evaluation. The authors wanted to ensure a baseline of security understanding, rather than to test for software security skill. However, participants who provided invalid data or completed the survey too quickly where excluded from evaluation.

As can be seen there is currently no common approach to detecting whether participants have programming skill. Each set of authors come up with their own ideas and no instrument has been tested in any rigorous manner.

\begin{table*} [h]\centering 
\caption{Overview of all questions}
\begin{adjustbox}{max width=\textwidth, max totalheight=\textheight} \begin{tabular}{clll}
\toprule
\textbf{No}&\textbf{Question}&\textbf{Abbreviation}&\textbf{Category}\\ 
\midrule
Q1&Which of these lesser-known programming languages
have you worked with before?&Unknown.Languages&Programming language recognition\\ 
\rowcolor{LightGray}Q2&Which of these websites do you most frequently use as
aid when programming?&Source.Usage&Information Sources\\ 
Q3&Choose the answer that best fits the description of a
compiler’s function.&Compiler&Basic Knowledge\\
\rowcolor{LightGray}Q4&Choose the answer that best fits the definition of a recursive function.&Recursive&Basic Knowledge\\ 
Q5&Choose the answer that best fits the description of an
algorithm.&Algorithm&Basic Knowledge\\ 
\rowcolor{LightGray}Q6&Which of these values would be the most fitting for a
Boolean?&Boolean&Basic Knowledge\\ 
Q7&Please pick all powers of 2.&Power.of.2&Basic Knowledge\\ 
\rowcolor{LightGray}Q8&Please translate the following binary number into a decimal number 101. &Bin.Conv&Number Formats\\ 
Q9&Please select all even binary numbers.&Bin.Even&Number Formats\\ 
\rowcolor{LightGray}Q10&Please select all valid hexadecimal numbers.&Hexa.Num&Number Formats\\ 
Q11&When multiplying two large numbers, your program
unexpectedly returns a negative number. What might
have caused this? &Error.Overflow& Finding Errors \\
\rowcolor{LightGray}Q12&What is the run time of the following code?&Runtime&Algorithmic runtime\\ 

Q13&When running the code, you get an error message for
line 6: Array index out of range. What would you change
to fix the problem?&Error.OutOfBound&Finding Errors\\ 
\rowcolor{LightGray}Q14&What is the purpose of the algorithm?&Sorting.Array&Code Comprehension\\ 

Q15&What is the parameter of the function?&Function.Param&Basic Knowledge\\ 
\rowcolor{LightGray}Q16&Please select the returned value of the pseudo code.&Backward.Loop&Code Comprehension\\ 

\bottomrule
\end{tabular} \end{adjustbox}  \label{tab:questions_overview} \end{table*} 

\section{Methodology}
In this section, we present the questions we designed to identify participants with programming skill and the studies we ran to evaluate it.

\subsection{Instrument Requirements}
Unlike Bergersen et al.'s~\cite{bergersen2014construction} instrument to determine programming skill which takes two days to complete, our goal is to assess whether participants have programming skill or not with as little effort as possible. For the questions to be used in pre-screening surveys or as quality control questions, we must ensure that they take people with programming skill only a few minutes at most to answer. Therefore, our requirements regarding the instrument were as follows:
\begin{itemize}
   \item \textbf{Effectiveness}: The instrument should be able to differentiate between programmers and non-programmers. Hence, the questions need to rely on domain knowledge and be complex enough so that only programmers can answer in a reasonable amount of time. It should not leave any scope for mere guesses.
    \item \textbf{Efficiency}: The instrument should consume as little time as possible. So, the goal is to frame questions that programmers can answer quickly. It is also desirable if it would help the participants without programming skill to decide quickly that they cannot answer the question, since we do not want to waste their time either.  
    \item \textbf{Robustness against cheating}:
    The instrument should be designed in a way that it becomes difficult for participants without programming skill to come by the answers, for instance, by using online search engines or forums on which Clickworkers exchange information about studies. 
    \item \textbf{Language independence}:
    The instrument should work regardless of the programming language the participants are skilled in. While it might also be useful to filter based on specific languages, that is beyond the scope of this paper. 
\end{itemize}

\subsection{Survey}
We used Dillman's pre-testing process to develop an online survey with 16 different 
questions~\cite[p.140-147]{dillman2011mail}. The Dillman's pre-testing process is a three-step approach to design survey questions in general. It includes literature review, using a think-aloud approach and conducting a pilot study. First, we reviewed related work and decided on different question types. We started with a larger pool of questions from related work. Since these questions rarely met all our requirements (e.g., time and effort), three researchers from Computer Science in different positions (graduate student, PhD student, professor) developed further questions by considering related work, but also examining the main concepts of programming. 
Second, another researcher went through the questions using a think-aloud approach. Third, we conducted a pilot study with two participants.
A summary of our questions can be found in Table~\ref{tab:questions_overview}. 
The full questions can be found in the supplementary material. 
A total of 16 questions were created and put under a number of categories: 
\begin{itemize}
    \item \textbf{Programming language recognition (Q1):} Q1 is a two part question. The first part asks the participants to self report their programming language skill for ``well-known programming languages'' such as C, C++, or Java. This part does not need to be evaluated for the instrument. The second part asks the participants to self-report their programming language skill for ``lesser-known programming languages'' such as Torg or Yod. However, all but one of the lesser known programming languages are fake. Both the parts offer a ``none of the above'' option. The idea behind this question is to see how quickly people with programming skill select their languages from the first list and then realize that they do not know any from the second. Our hypothesis is that they will probably also realize that most of the names on the second list are fake and select ``none of the above,'' while non-programmers who want to falsely claim skill would select a couple. 
    \item \textbf{Information sources (Q2)}: Previous studies have analyzed what kind of websites developers use while programming, with the most popular being Stack Overflow~\cite{Fisher17}. Hence, Q2 contains Stack Overflow and some decoy options. Participants with programming skill can quickly pick Stack Overflow, while non-programmers might not be aware at all.
    \item \textbf{Basic knowledge (Q3-Q7, Q15)}: These questions cover general programming knowledge, for instance, regarding what a compiler does, what an algorithm is, and what a recursive function is.
    \item \textbf{Number formats (Q8-Q10)}: We asked simple questions related to hexadecimal and binary conversion. Most computer science programs or programming language tutorials deal with binary and hexadecimal numbers; hence, most non-programmers will not have much experience with this.
    \item \textbf{Finding errors (Q11, Q13)}: A good test of programming skill would have been asking the participants to actually write some code; however, that would cost more time than can be permitted and is hard to automatically verify. Therefore, as an alternative, we asked the participants to find errors in code snippets or explain why the errors occurred. 
    \item \textbf{Algorithmic runtime (Q12):} We also added a basic question about the run time of some simple pseudocode.
    \item \textbf{Program-comprehension (Q14, Q16)}: We showed the participants two pseudo-algorithms and asked them about their functionality and output. It is important to note that Q16 was taken from the study by Danilova et al.~\cite{danilova2020one}.
\end{itemize}

The questions Q13+Q14 as well as Q15+Q16 are based on the same pseudocode. Hereafter, we will refer to each pair of these questions as a ``question block.''
While we tried to keep the time spent on each question short, especially for participants with programming skill, we also had to enable automatic evaluation and make the questions robust against random guesses.
Thus, we opted for closed multiple choice questions with 5 to 6 possible answers for each question.
Most questions had one unambiguous correct answer. Exceptions were 2 of the number-format questions, where participants were asked to select all correct solutions. The incorrect answers were chosen in a way as to look plausible to participants without programming skill.

Furthermore, an attention check question~\cite{kung2018attention} appeared randomly during the survey to filter out careless respondents. For the attention check question---This is an attention check question. Please select the answer ``Octal"---the correct item needed to be picked.
The questions and answer options were shown in a randomized order to mitigate response fatigue and response order effects~\cite{lavrakas2008encyclopedia}. 

We used 2 versions of this survey. The initial version included “I don't know” or “I don't program” answer options. We included these options because we wanted to minimize guessing at this stage so we could get an accurate view of what non-programmers state did not know. 
The second version of the survey was conducted in an adversarial setting (see Section~\ref{attackscenario}). Here, the participants were given a monetary reward for each correct answer and the “I don't know” or “I don't program” options were removed. This setting simulated a screening setting in which non-programmers might try to guess the correct answers to take part in a well-compensated survey. 

After completing the programming questions, the participants were asked to answer demographic questions, including ones related to their age, job, and programming experience. 
The full questionnaire can be found in the supplementary material. 
For evaluation and for testing our time requirement, we set a timer for each question to measure how much time was spent to solve it. 


\subsection{Statistical Testing}
We categorized the answers as \textit{correct} or \textit{incorrect}.
In order to test whether the different groups had different success rates for different questions, we used the Fisher's exact test (FET)~\cite[p.~816]{field2012discovering} on each question. We reported confidence intervals (CI) and odds ratio (OR) to interpret the details of the tests. We corrected all Fisher's exact tests for multiple testing using the Bonferroni-Holm correction. 


To analyze the entire set of questions, we used latent class analysis, as it is suited the categorical data~\cite{mccutcheon1987latent,collins2009latent,linzer2011polca}. The latent class analysis reveals whether the data shows a number of distinct classes. Our assumption was that we will get two: participants with and without programming skill. We tested the models with more classes as well but selected the one with the lowest Bayesian information
criterion (BIC).


\begin{table*}[h] \centering\footnotesize
\caption{Demographics of the participants (n = 249)}
\begin{adjustbox}{max width = \textwidth} 
\begin{threeparttable}

\begin{tabular}{llc|llp{6cm}p{6.5cm}}

\toprule 
\textbf{Group}&\textbf{Sample}&\textbf{n}&\textbf{Gender}&\textbf{Age}&\textbf{Country of Residence}&\textbf{General Programming Experience [years]}\\
\midrule\footnotesize
\multirow{2}{*}{\textbf{Programmer}}&\textit{CS students }&17&Female:  4, male:  13&min:  19, max:  30, mean: 21.82, md:  20, sd:  3.26&Germany:  17&min:   2,  max:   16,  mean: 5.31,  md:   4.5,  sd:   3.41, NA:  1\\
&\textit{Professional developers}&33&Female:  2, male:  31&min:  25, max:  55, mean: 36.45, md:  36, sd:  8.04&Germany:  31, Austria:  2&min:   2,  max:   30,  mean: 13.09, md:  15, sd:  7.31\\\midrule
\multirow{2}{*}{\textbf{Non-Programmer}}&\textit{Econ students}&50&Female:    36,   male:    13, PNTA:  1& min:  18, max:  28, mean: 22.66, md:  23, sd:  2.3&Germany:  49, NA:  1&min:   0,  max:   2,  mean: 0.21, md:  0, sd:  0.52\\
&\textit{Clickworkers without programming skill}&50&Female:    20,   male:    29, PNTA:  1&min:  19, max: 67, mean:   34.02,  md:   31.5, sd: 10.43&Germany:    21,   UK:    8, USA:   6,  Other:   15 
&min:   0,  max:   25,  mean: 1.63, md:  0.25, sd:  4.1\\\midrule
\textbf{Test Group}&\textit{Clickworkers with programming skill}&52&Female:    10,   male:    41, PNTA:  1&min:  18, max:  58, mean: 33.73, md:  31.5, sd:  9.81&Germany:    22,   UK:    5, Other:   25 
&min:   0,  max:   30,  mean: 6.08,   md:    3,   sd:    7.78, NA:  1\\\midrule
\textbf{Attack Scenario}&\textit{Clickworkers without programming skill}&47&Female:    18,   male:    28, PNTA:  1&min:  19, max:  54, mean: 32.15, md:  30, sd:  9.15&Germany:    16,   Italy:    4, Spain:  4, USA:  4,  Other:  19    
&min:   0,  max:   26,  mean: 2.22, md:  1, sd:  4.82\\

\bottomrule
\end{tabular}

\begin{tablenotes}
\centering
\item \textbf{md:} median. \textbf{PNTA:} prefer not to answer. See supplementary material 
for further details on occupation and country of residence.
\end{tablenotes}

\end{threeparttable}
\end{adjustbox}\label{tab:  demographics2} 
\end{table*} 

\subsection{Participants}
\subsubsection*{Recruitment}
We sampled through different channels to obtain data from different groups.
We sampled 17 CS students using the mailing-list of an advanced programming lecture from the undergraduate program of our university. As compensation, the students received bonus points for their exam admission. All the CS students passed our attention check. 

Additionally, we invited 49 professional developers from personal contacts and from a database of professional developers who took part in our past programming studies and agreed to take part in future studies. Thirty-five participants completed the survey. We excluded one from our data set because we were not able to identify the participant on our invitation list, and it seemed like that the study link was forwarded. All the professionals passed our attention check and received 10 euros for their participation. 
We combined these two groups to form our ground truth since we knew for sure that they all have programming skill.

Next, we recruited 54 students in cooperation with the behavioral economics group from our university. They have a recruitment system which sends email invitations for studies to enrolled users. 
The majority of them were economics students; however, others potentially including computer science students could have enrolled as well. 
We refer to these participants as \emph{econ students}.
Of these 54 participants, 50 passed the attention check. Based on the question of self-reported programming experience, 10 participants had at least some (0.5 years) experience. 
The participants received 5 euros as compensation.

We also recruited 75 participants from Clickworker, who did not have any programming skill.
Of these 75 participants, 53 passed the attention check and 50 completed the survey. They received 2.50 euros for their participation.\footnote{This fulfills the minimum wage requirement. Our compensation was higher than the recommendation of the platform, which was 1.50 euros for 10 minutes.} However, in contrast to all other samples, we used the default Clickworker invitation description which cautions Clickworker participants that attention check questions needed to be solved correctly to receive the payment. We accepted this difference since this is the norm on Clickworker;  it is the norm to pay participants even if they fail the attention check questions on the other recruiting platforms.

We combined the last two groups as our ground for participants without programming skill. However, the situation is not as clear cut as with the programmers above since CS students can also be enrolled in the econ platform, and both econ students and regular Clickworker participants might have programming skill even though they do not state it. In this combined group we have 35 out of 100 participants who stated that they have programming experience. However, since we have no way of verifying this properly, we did not remove them from our evaluation, to be on the conservative side. 

Further, we recruited 55 Clickworker participants who listed programming as a skill in their profile. The hiring conditions were the same as above. Of these 55 participants, 52 passed the attention check. While the above-mentioned groups were used to design our screening instrument, we used this last group as a real-life test to see how many of these would pass our screening questions. 

Finally, we tested the 8 best questions using an attack scenario with 51 participants from Clickworker, of whom 47 passed the attention check. We paid a base compensation of 2 euros to Clickworkers who did not state that they had programming skill and provided a bonus payment of 2 euros for each of the 8 correct answers. This should simulate a non-programmer adversary who wants to pass a screener question to be able to take part in a well-compensated developer study.



\subsubsection*{Demographics}
The demographics of our tested groups can be found in Table~\ref{tab: demographics2}. 
Of all the 249 participants, 155 were male, 90 were female, and 4 preferred not to answer the question.
From the programmer group and the test group, almost all participants were male (developer: 44/50 male; test group: 41/52 male). By contrast, from the non-programmer group, more participants were female (out of 100: 56 female, 42 male, and 2 preferred not to tell). 
The majority of the participants were from Germany.

While professional developers reported to have on average 13.09 years of programming experience (min: 2, max: 30, median (md): 15, standard deviation (sd): 7.3), CS students indicated, on average, 5.31 years of experience (min: 2, max: 16, md: 4.5, sd: 3.41). All the participants from the programmer group indicated to have worked with Java before. A total of 31 of the 50 programmer participants indicated to have worked with JavaScript, 34 with C, 30 with Python, 28 with C++, 22 with PHP, 20 with C\#, 19 with Shell, 8 with Typescript, 3 with Ruby, another 3 with Groovy, and 2 with Go.

Clickworker participants who claimed to have programming skill in their profile indicated in our survey to have, on average, 6.08 years of experience (min: 0, max: 30, md: 3, sd: 7.78). By contrast, Clickworker participants who did not indicate to have programming skill in their profile reported in the survey to have, on average, 1.63 years of experience (min: 0, max: 25, md: 0.25, sd: 4.10). Finally, most of the econ students reported not to have programming experience at all (mean: 0.21, min: 0, max: 2, sd: 0.52).


\section{Ethics}
The institutional review board of our university approved our project. The participants of our study were provided with a consent form outlining the scope of the study and the data use and retention policies, and we also complied with the General Data Protection Regulation (GDPR). The participants were informed of the practices used to process and store their data, and that they could withdraw their data during or after the study without any consequences. The participants were asked to download the consent form for their own use and information.

\section{Limitations}
We compensated each sample differently, as the different groups had different payment expectations. However, it could be that the different compensation levels affected the results. We found a couple of participants in the non-programmer group who looked like they had programming skill. There might also have been some with programming skill whom we did not recognize. However, we think that the number of programmers that accidentally fell in the non-programmer group was not significant enough to interfere with our study. If anything, our instrument’s performance will be under-reported since we count any unknown programmer in the non-programmer group who is identified as a programmer as a failure for our instrument. 
While we have recruited a mix of CS students, econ students, professional developers, and Clickworkers with and without programming skill, we do not claim that this is representative for all programmers. Hence, further studies will be needed to extend and validate our results.

\section{Results}
In this section, we describe the effectiveness and efficiency of our 16 defined screener questions.
However, the questions should also be effective in a way that that the number of programmers who fail and the number of non-programmers who either know or guess correctly should be low.
Therefore, we additionally tested our questions with a test group to reduce our question set to the most promising questions and evaluated them in an adversarial scenario.

\subsection{Effectiveness}
Except for Q1, all Fisher's exact tests (15/16) were highly significant even after the Bonferroni-Holm correction. This, in turn, indicates that the remaining 15 questions were different in the distributions of correct and incorrect answers between the non-programmer and programmer groups. 

\begin{figure}[]
\centering
    \includegraphics[width=0.5\textwidth]{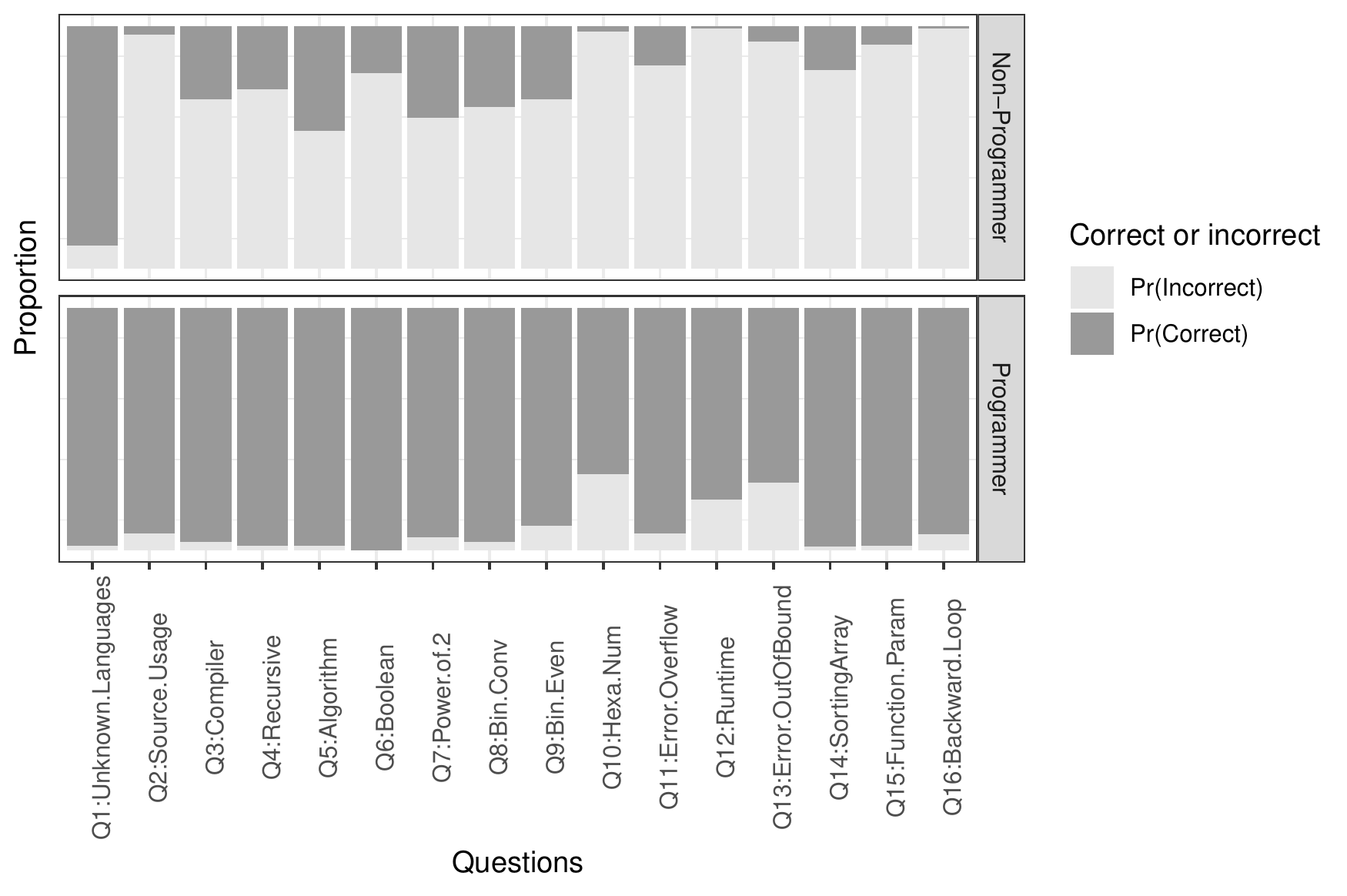}
        \caption{Plot of latent class analysis with the programmer and non-programmer groups.}
        \label{fig:LCM}
\end{figure}

We conducted a latent class analysis on the 16 questions. 
We chose a model with two groups ($G^{2}(2)$: 613.02 (Likelihood ratio/deviance statistic) 
$\chi^{2}$(2): 174111.9 (Chi-square goodness of fit), maximum log-likelihood: -907.89, entropy: 6.06), since the BIC was lower for models with more classes and it fit well to our assumption. Figure \ref{fig:LCM} visualizes the proportion of probabilities for choosing the correct or incorrect items according to the groups. The predicted class shares were 0.62 for Class 1 (non-programmer) and 0.38 for Class 2 (programmer). Thus, according to the class analysis, two classes can be distinguished within this sample. This fits into our self-chosen samples, since we had 50 programmers and 100 non-programmers sampled. Indeed, our questions are applicable to split a population according to the answers the participants provided. 
In the following, we describe the effectiveness of each question in detail.

\subsubsection*{Programming Language Recognition (Q1)}
We found that almost all the non-programmers (91\% or 91 of 100) and programmers (98\% or 49 of 50) answered the question for lesser-known languages correctly. That means in this case, they chose the answer ``None of the above.'' 
Nine participants from the non-programmer group and one participant from the programmer group selected non-existent programming languages. This shows that without incentive most non-programming participants had no reason to pick fake languages, although it is interesting that roughly 10\% of non-programmers selected some of the fake languages. 

\subsubsection*{Information Sources (Q2)}
All the 50 programmers (100\%) selected Stack Overflow as one of the most used aids for programming. No other source was reported.
Most participants from the non-programming group reported that they do not program (60\% or 60/100), while 15  said that they do not use any of the listed websites in the questionnaire for programming. Interestingly, 9 participants chose MemoryAlpha, 8 selected Wikipedia, 2 other participants marked LinkedIn and only 6 participants from the non-programmer group picked Stack Overflow as their answers.

\subsubsection*{Basic Knowledge (Q3 to Q7 and Q15)}
First, we analyzed the answers given on the description of a compiler (Q3).
All the participants from the programmer group and 33\% participants from the non-programmer group chose the correct answer. As stated earlier, our non-programmer group is not as controlled as our programmer group because 35 participants from the group stated that they had some programming skill. However, of the 33 correct answers only 23 came from this group, i.e., 12 participants who stated to have programming skill did not give the right answer for Q3, whereas 10 who stated that they had no skill got it right. This similar pattern that emerged was found in all the knowledge questions with no clear distinction visible between the two subgroups. Thus, for simplicity we will not report further on this, but a full overview can be found in the supplementary material. 

Second, we analyzed the answers given on the definition of a recursive function (Q4).
The answers were very similar to the compiler question. All programmers could answer the question correctly, whereas 30\% of the non-programmers got the correct answer.
The performance for Q5 that was about the definition of an algorithm (Q5) was even worse, as 47 (out of 100) were able to answer it correctly.
One programmer failed the question. 
The effect on the boolean question (Q6) was significant as well,
since all the programmers successfully chose the correct answer, while only 25\% of the non-programmers answered correctly.

We also included a question (Q7) about the power of two, since we thought most programmers work with powers of two. The answers were only marked as correct if all the powers of two were selected. Two programmers answered incorrectly, while the rest chose the correct answer.
From the non-programmer group, 41\% managed to pick the right powers of two as well.
Furthermore, we asked our participants to pick what a function's parameter is (Q15).
This question was answered correctly by all the programmers, while only 13 participants from the non-programmer group selected the correct answer.
It seems that programmers were familiar with the definition, while the non-programmers struggled to answer this question correctly. 

\subsubsection*{Number Formats (Q8 to Q10)}
First, we asked our participants to convert a number from the binary system into the decimal system (Q8). Forty-eight of the 50 programmers and 38 of 100 non-programmers solved the question correctly. 
With regard to picking all the even binary numbers question (Q9) with multiple answers, 5 programmers failed to select all the correct answers. However, 34 out of 100 participants from the non-programmer group got the correct answer. 
To test the participants' knowledge with hexadecimal numbers, we asked them to select all the valid hexadecimal numbers in Q10. To get the correct answer, multiple choices needed to be selected. While the effect of groups on the answers was again significant,
this question seemed to be very challenging for all the participants. 30\% (15 of 50) of the programmers were unable to solve this question. From the non-programmer group, only 6 participants succeeded in selecting the correct answers. 

\subsubsection*{Finding Errors (Q11 and Q13)}
Fixing bugs takes a large percentage of a programmer's working time. With regard to the question of what could happen if two large numbers are multiplied and a negative number is returned (Q11), the difference of the two groups was significant.
Forty-seven of the 50 programmers were able to answer this question correctly as well as 21 of 100 non-programmers.
We also investigated the common errors that programmers face during programming and requested them to solve an ErrorOutOfBound (Q13). 
Only a few non-programmers answered the question correctly (9 of 100). However, the programmers also seemed to have trouble with this question, because ``only'' 38 out of 50 selected the correct answer.

\subsubsection*{Algorithmic Runtime (Q12)}
The question for the run-time seemed to be very difficult for both the groups. We concluded that even the participants with programming skill were not very familiar with algorithmic run-times. The effect of the group variable was significant because the proportion of correct answers differed between both the groups. 

\subsubsection*{Program-comprehension (Q14 and Q16)}
Comprehension of the sorting array pseudocode seemed to be an easier task for the programmers, because almost all of them answered Q14 correctly (49 of 50). 
From the non-programmer group, 24 selected the correct answer.
Furthermore, 
3 programmers were unable to solve the ``hello world'' pseudocode task (Q16) correctly. Additionally, only 7\% of non-programmers choose the correct answer. Consequently, the difference in correct answers between both the groups was significant.

\subsection{Efficiency}
The participants in the non-programmer group recorded 7.87 minutes as the median time to complete the whole questionnaire, 
while the participants in the programmer group finished the survey with a median time of 10.87 minutes. 
All questions showed a mean under 100 seconds; thus, all questions fulfilled our efficiency requirement. We found that answering knowledge questions took the least time, while answering questions about number conversion took longer.  
As expected, each of the two question blocks including pseudocode (Q13+Q14 and Q15+Q16) took more time as compared to other multiple choice questions. We found that participants with programming skill needed more time to answer them. The reasons for this could be that non-programmers selected ``I don't know" or gave a random answer quickly, since they knew they could not understand the code. The adversarial measurements in Section \ref{attackscenario} are more relevant for these questions.  

A visualization of the mean times for both programmer and non-programmer groups according to each task block, an overview for the number of correct and incorrect answers of the programmer and non-programmer groups for each question as well as the statistical analysis summary are available in the supplementary material. 

\section{Testing the Instrument}
We tested the instrument in two scenarios: (1) Non-Adversarial Test group and (2) Adversarial Attack group. First, we tested our survey instrument with a set of 52 participants from Clickworker who indicated in their profiles to have programming skill. This scenario is close to how real studies would be conducted, i.e., researchers use a platform like Clickworker to recruit participants who state that they have programming skill. We did not offer significantly higher compensation to minimize the incentive to claim skill to participate in our survey. We also included in the ``I don't know'' options, since we wanted to see how many of the self-reported programmers from Clickworker would choose that they did not know answers in a non-adversarial setting. 
The results also served for comparison of self-reported programming skill of the participants from Clickworker with professional developers from our controlled sample.
Later, we selected the most promising questions and tested them in an adversarial scenario, where we recruited 47 participants from Clickworker who did not state to have programming skill. In the introduction, we explained the goal of our study and asked the participants to try and pass themselves off as programmers. To incentivize them, we paid a base fee of 2 euro and an additional 2 euro for every correct answer for the 8 questions. In this scenario, we removed the 
``I don't know'' options because we needed to evaluate the questions in an adversarial setting and measure their guessability. This scenario simulates people without any programming skill trying to break our screener questions to take part in a well compensated developer study. 

\subsection{Non-adversarial Test Group}
\textbf{Programming language skill (Q1):}
Forty-eight of the 52 participants did not select any imaginary languages when asked for skill with lesser-known programming languages. The majority (46) selected ``None of the above" and 2 selected ``SHROUD.''

\textbf{Information sources (Q2):}
For answering the question for the most used information sources, 30 of the 52 participants selected Stack Overflow, while the rest chose that they did not program or have not used any of the websites suggested (9 of 52). Five chose Wikipedia. 

\textbf{Basic knowledge (Q3 to Q7 and Q15):}
Forty-one of the 52 participants correctly selected the description of a compiler (Q3), whereas 11 failed. For the description of a recursive function (Q4) and the value of a boolean (Q6), 41 selected the correct answer. For the description of an algorithm (Q5), only 5 failed to select the correct answer. Forty participants were able to select all the powers of two (Q7) while 10 did not. Thirty-three participants selected the function's parameter (Q15) correctly.

\textbf{Number Formats (Q8 to Q10):}
Forty participants were able to select the correct answer for a simple binary conversion (Q8), while 6 selected a wrong answer and another 6 reported to not know the answer.
The multiple response questions seemed to be more difficult.
Thirty-two participants could correctly select all the even binary numbers (Q9).  
Further, only 30 participants selected all the valid hexadecimal numbers (Q10), while 17 failed or selected ``I don't know'' (5). 

\textbf{Error (Q11 and Q13):}
Thirty-five participants correctly selected an overflow as source of error in Q11, and 12 participants reported that they did not know the answer and the rest selected an underflow.
For the array out of bound error (Q13), only 17 of 52 (32\%) participants were able to select the correct solution. 


\textbf{Algorithmic runtime (Q12):}
The runtime question seemed to be the hardest one for the participants from the test set. Only 15 answered correctly, 21 did not know the answer and the rest selected a wrong response. 

\textbf{Program-comprehension (Q14 and Q16):}
Thirty-one participants selected the correct purpose of our sorting array pseudocode (Q14). 
Interestingly, only 24 participants selected the correct output of the hello world pseudocode (Q16), whereas 13 
participants selected the wrong answer ``hello world."

\subsubsection*{Comparison of the Groups}
\begin{figure}[]
\centering
    \includegraphics[width=0.4\textwidth]{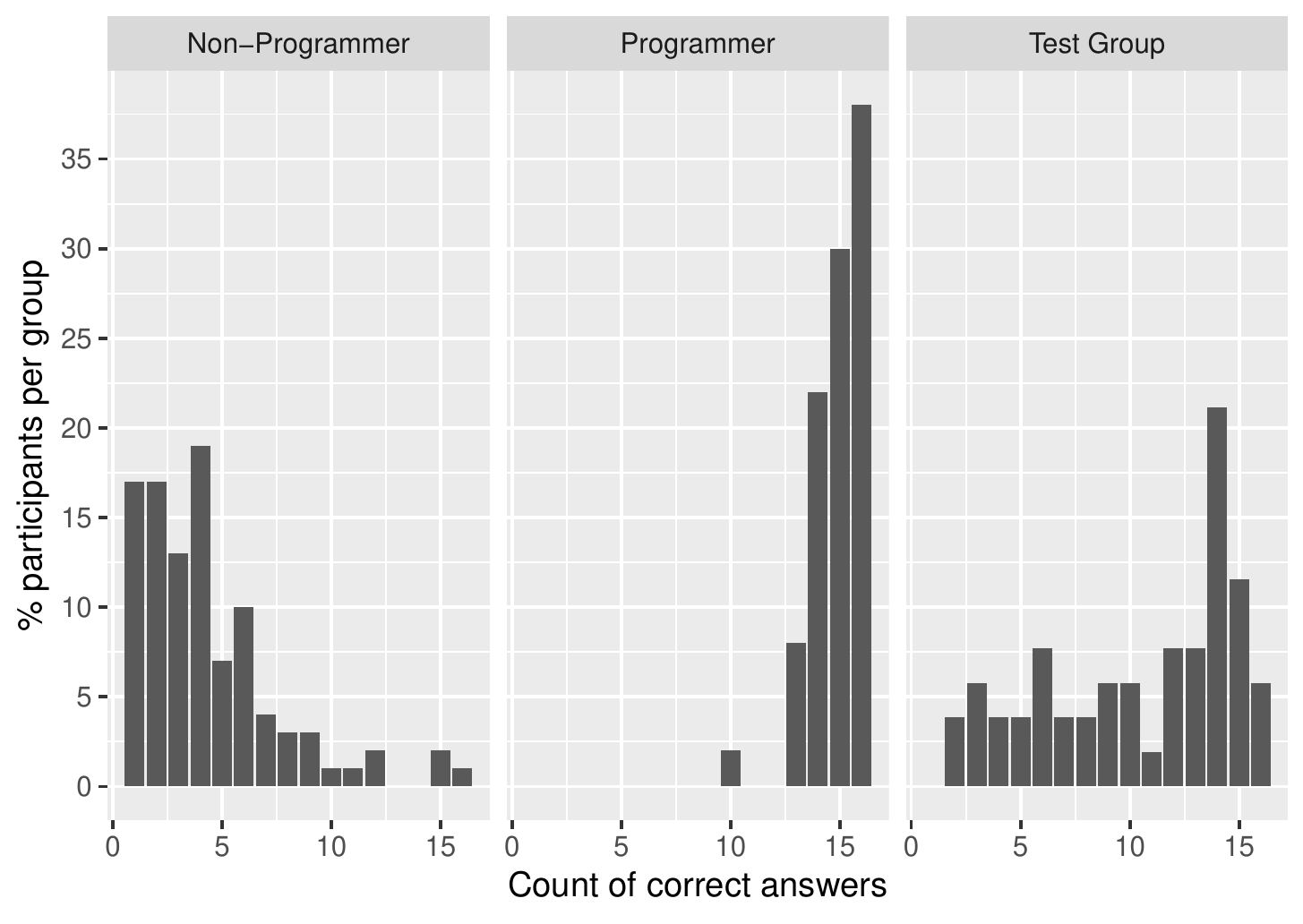}
        \caption{Number of correct solutions of all 16 questions per group.} 
        \label{fig:sumsolution}
\end{figure}

Table \ref{table:sortedcomparison} shows the correct answer rates of the programmer, non-programmer, and the non-adversarial test group for the 16 questions. The test group (Clickworker who stated that they had programming skill) always had more correct answers than the non-programmer group (econ students and non-programmer Clickworkers) but less than the programmer group (CS students and company developers). This suggests that some Clickworker participants do not interpret programming skill the way we do or falsely indicated in their profile to have programming skill. In either case, it is important to be able to identify these participants during screening. Interestingly, 7 participants from this test group selected ``\textit{I don't program},'' when we asked for their level of programming expertise at the end of the survey, despite listing programming as a skill in their profile. The number of correct solutions for all the 16 questions per group is visualized in Figure~\ref{fig:sumsolution}.



\begin{table} []\centering 
\begin{threeparttable}[!h] 
\footnotesize 
\caption{Percentages of correct answers in each group.}

\begin{tabular}{cp{2cm}ccc}

\toprule

\multirow{2}{*}{\textbf{No}}&\multirow{2}{*}{\textbf{Question}}&\multirow{2}{*}{\textbf{Programmers}}&\textbf{Non-}&\textbf{Test}\\
&&&\textbf{Programmers}&\textbf{Group}\\
\midrule
\rowcolor{mine}Q2&Source.Usage&100\%&6\%&57.69\%\\ 
\rowcolor{mine}Q15&Function.Param&100\%&13\%&63.46\%\\ 
\rowcolor{mine}Q6&Boolean&100\%&25\%&78.84\%\\ 
\rowcolor{mine}Q4&Recursive&100\%&30\%&78.84\%\\
\rowcolor{mine}Q3&Compiler&100\%&33\%&78.84\%\\

\rowcolor{mine2}Q14&Sorting.Array&98\%&24\%&59.61\%\\ 
\rowcolor{mine2}Q5&Algorithm&98\%&47\%&90.38\%\\ 
\rowcolor{mine3}Q1&Unknown.Languages&98\%&91\%&96.15\%\\ 
\rowcolor{mine2}Q8&Bin.Conv&96\%&38\%&76.92\%\\ 
\rowcolor{mine2}Q7&Power.of.2&96\%&41\%&76.92\%\\ 
\rowcolor{mine2}Q16&Backward.Loop&94\%&7\%&46.15\%\\ 
\rowcolor{mine2}Q11&Error.Overflow&94\%&21\%&67.30\%\\ 

\rowcolor{mine2}Q9&Bin.Even&90\%&34\%&61.53\%\\ 
\rowcolor{mine3}Q12&Runtime&80\%&6\%&28.74\%\\ 
\rowcolor{mine3}Q13&Error.OutOfBound&76\%&9\%&32.69\%\\ 
\rowcolor{mine3}Q10&Hexa.Num&70\%&6\%&57.69\%\\

\bottomrule
\end{tabular} 
\begin{tablenotes}
\item \small{The table is sorted descended by the column ``Programmers'' and ascended by the column ``Non-Programmers,'' because the most promising questions require to be correctly answered by programmers and incorrectly by non-programmers.} 

\end{tablenotes}
\label{table:sortedcomparison} 
\end{threeparttable}
\end{table} 

\begin{table*}\centering 
\begin{threeparttable}[!h] 
\footnotesize
\caption{Overview of screening question recommendations for programming skill}
\centering \begin{tabular}{cccp{3cm}p{4cm}p{2cm}}
\toprule
\textbf{Question}&\textbf{Recommended} &\textbf{Suggested time limit [seconds]}& \textbf{Excluded programmers with time limit (n = 50)}&\textbf{Attackers (n = 47) (included$|$excluded)} \\
\rowcolor{mine3}Q1&	\textcolor{red}\xmark &- & -  & -\\
\rowcolor{mine2}Q2&\textcolor{darkgreen}	\cmark& 30&	2 & 10$|$19\\
\rowcolor{mine2}Q3&\textcolor{darkgreen}	\cmark&	60 &	3 &22$|$14 \\
\rowcolor{mine2}Q4&	\textcolor{darkgreen}\cmark	& 30&	4&8$|$28\\
\rowcolor{mine2}Q6&	\textcolor{darkgreen}\cmark&30&	0&14$|$22\\
\rowcolor{mine3}Q14&\textcolor{red}\xmark& - &-&-\\
\rowcolor{mine}Q15& \textcolor{darkgreen}\cmark& Not necessary &-&-\\
\rowcolor{mine}Q16& \textcolor{darkgreen}\cmark& Not necessary &-&-\\
\bottomrule
\end{tabular}
\begin{tablenotes}
\item \small{The table shows an overview of our recommendations for the eight questions tested in the attack scenario.  Colors: red = not recommended, green = recommended without restrictions, yellow = recommended but with time limit}\\ 
\end{tablenotes}
\label{table:recommendation}  
\end{threeparttable}
\end{table*}

\subsection{Attack Scenario}
\label{attackscenario}
Based on the above findings, we selected the most promising questions (Q2 to Q4, Q6, Q14, and Q15) by excluding all the questions that did not achieve a correct answer rate of at least 98\% from the programmer group. We also excluded all the questions where more than 40\% of the non-programmer group gave the correct answer (Table~\ref{table:sortedcomparison}). We chose these cut-offs for several reasons. First, we did not require 100\% correctness from the programmers because even programmers make mistakes. Additionally, we allowed some false positives from the non-programmers because we need to keep the questions quick and simple to not lose programmers to the screening process. Since, in most scenarios, we would recommend the use of multiple questions, the false-positive rate will be lower due to the combination. 
In addition to the 6 questions mentioned above, we included Q1 and Q16 to our set of most promising questions because we expected them to perform better in the attack scenario. 
After each question, we asked the participants whether they looked up the answer on the Internet or solved it on their own (see supplementary material for more details). 

\textbf{Programming language skill (Q1):}
In the adversarial setting, we tested Q1 in two versions: 1) both parts of the questions (real and fake programming languages) and 2) on its own (only fake programming languages). Both versions were ineffective with only 4 of 18 in version 1 and 3 of 29 in version 2 where the participants chose non-existent programming languages. Our assumption that non-programmers would pick these turned out to be false.

\textbf{Sources (Q2):}
Twenty-nine of the 47 
Clickworkers (61\%) reported ``Stack Overflow" to be their most used information source for programming. Six chose "None of the above," while 7 chose ``Wikipedia" and 5 ``MemoryAlpha." 

\textbf{Compiler's function (Q3):}
Thirty-six of the 47 participants (76\%) chose the correct answer for the functionality of a compiler. 

\textbf{Recursive function (Q4) and boolean (Q6):}
76\% (36 of 47) participants correctly defined a recursive function 
and answered the value that a boolean can take.

\textbf{Sorting array (Q14):}
Twenty-nine of the 47 participants (62\%) answered the sorting algorithm question correctly. 
This might be the first indication of algorithms being more difficult for the participants without programming skill to look up. 

\textbf{Parameter of a function (Q15):}
Thirteen of the 47 participants (27\%) selected the correct answer. 

\textbf{Hello World (Q16):}
25\% of Clickworkers (12 of 47) 
picked the correct answer for the hello world question.

\begin{figure}[]
\centering
    \includegraphics[width=0.5\textwidth]{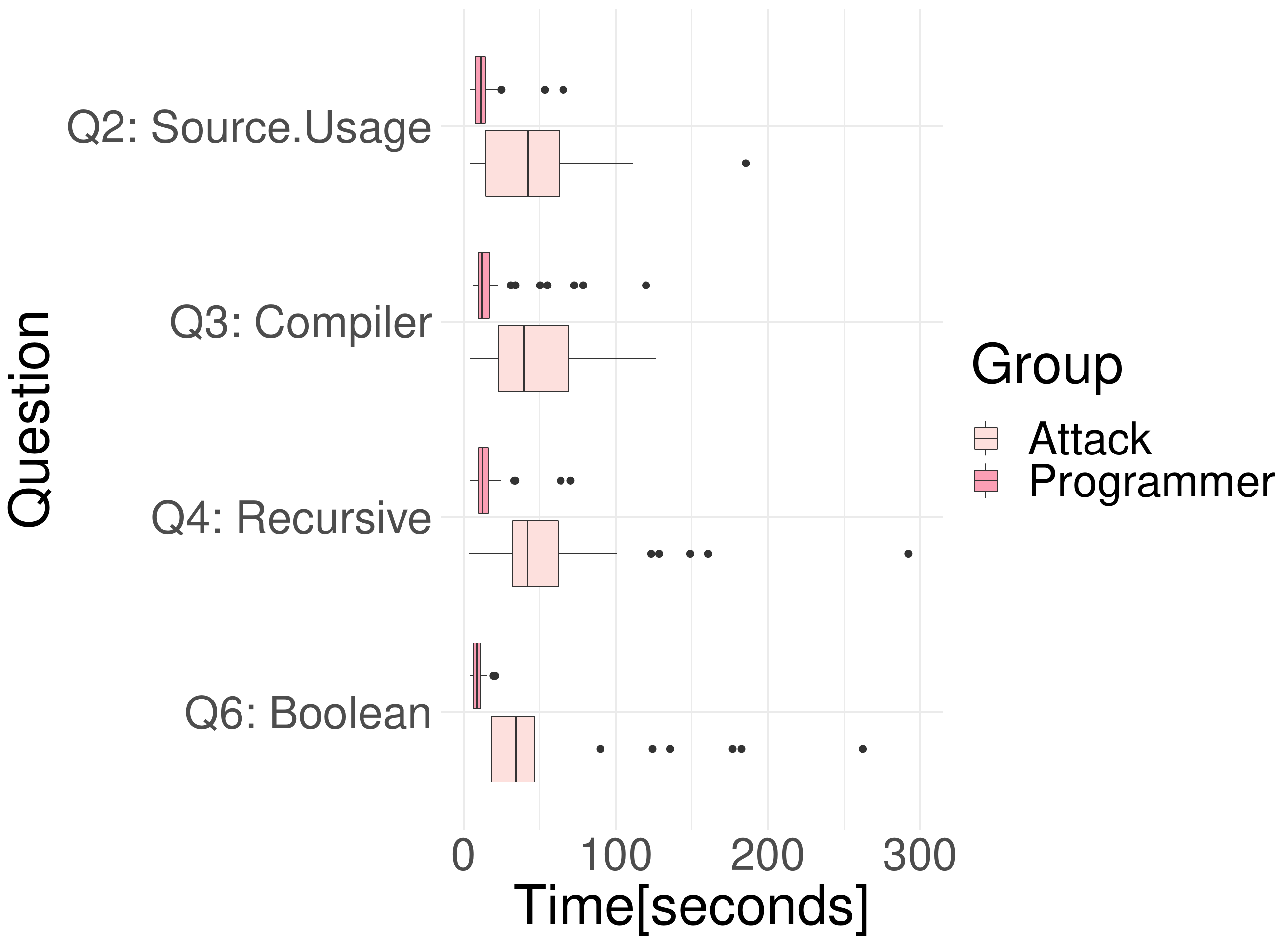}
        \caption{Comparison knowledge questions: Time to solve each question \textbf{correctly}.}
        \label{fig:attack_time}
\end{figure}

Our analysis showed that Q15 and Q16 performed best according to the correct answer rates in the attack scenario. 
Figure \ref{fig:attack_time} shows the time taken to solve the questions correctly for the programmer and the attack groups. We excluded Q1 since we tested 2 versions of this question, as described above, and the question performed so poorly that it was not considered. 
We also did not directly compare Q14, Q15, and Q16 to the original questions since, in the prior analysis, they were part of a question block. In the attack scenario, we split them in order to get the information for each question whether participants googled the answers or not. All knowledge questions, except Q2 ($p$ = 0.054), showed a significant time difference (wilcoxon rank sum test) between the developer and attacker groups, with the attackers taking significantly longer. 





\section{Discussion and Recommendations}
When conducting our non-adversarial evaluation with the Clickworker test group, we only selected participants whose profiles included programming skill. The results of this test group demonstrated that for the kind of developer studies that are common in our community, it is not recommendable to rely on the self-reported programming skill or a platform’s recruitment features. In our test set, 42\% of the Clickworker programmers got fewer correct answers than the poorest performers in our ground truth programmers group. They mostly got all the answers right while many Clickworker developers got less than half right answers. Considering the small sample sizes, which are common in developer studies, even a small amount of noise can mask true effects or worse create false effects. Having potentially one-third participants without programming skill in a developer survey can cause significant disruption, and we highly recommend using screening questions to avoid this. The fact that we removed 10 of our 16 questions, since they proved to be less effective than we had hoped, suggests that the current practice of some researchers creating and using untested screener questions is sub-optimal. We hope that our tested questions can be a first step toward creating a common screener instrument for our community.  

\begin{figure}[]
\centering
    \includegraphics[width=0.4\textwidth]{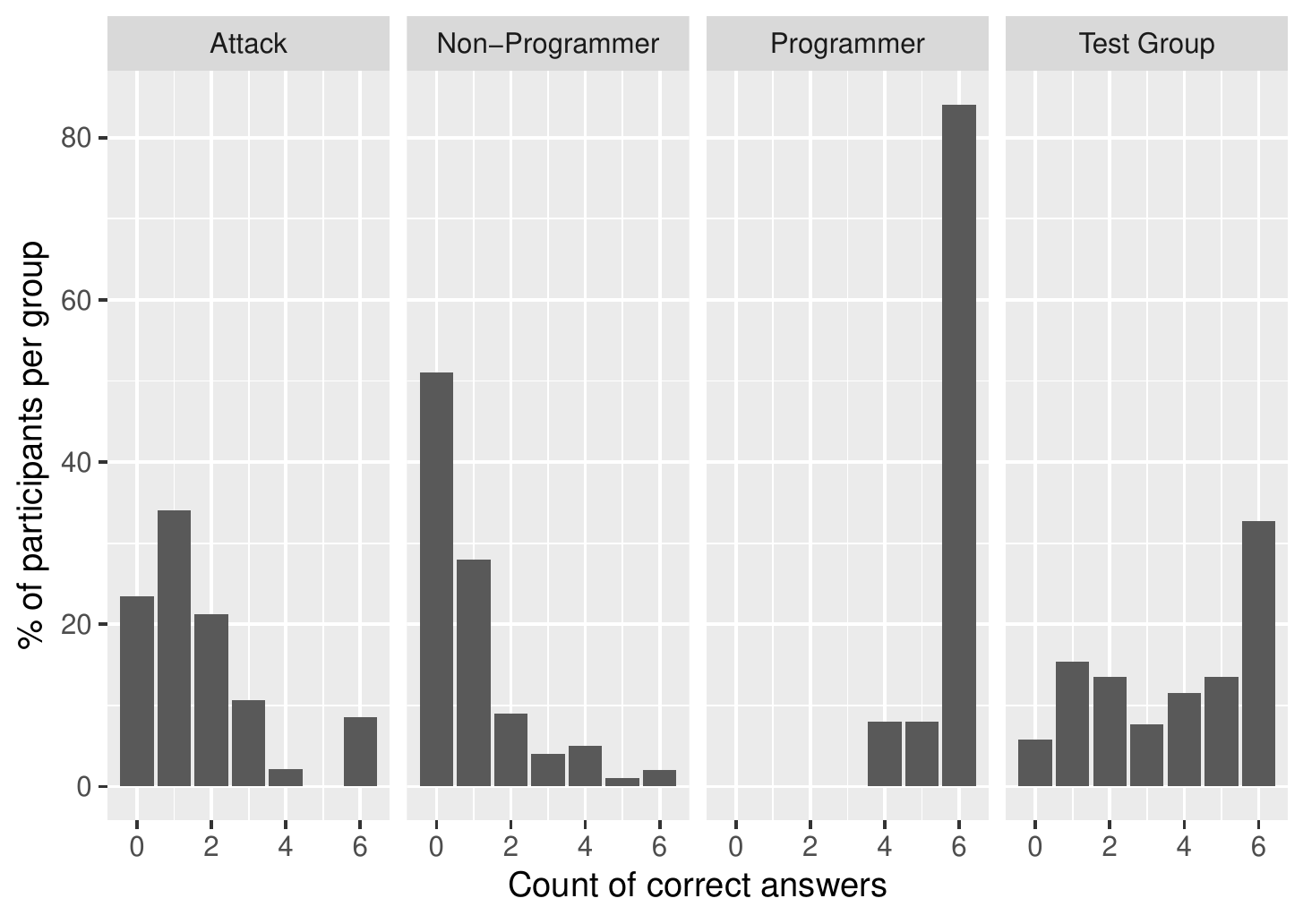}
        \caption{Distribution of correct solutions of all 6 recommended questions per group with time limits applied as in Table \ref{table:recommendation}. }
        \label{fig:recommendHist}
\end{figure}

We recommend the use of screener questions for studies targeting people with programming skill, especially if these studies do not contain a programming task. 
Table~\ref{table:recommendation} summarizes our screener question recommendations.
The most effective but also the slowest questions were the code comprehension questions, i.e., Q15 and Q16. If time penalty is acceptable, we recommend using these or similar questions as screeners. If this is not feasible, we recommend to randomly use one of the four knowledge questions (Q2 to Q4 and Q6) with a time limit, since our attack group demonstrated that these question can be looked up. The time limit can be used to configure the false rejection participants with programming skill and the false acceptance of those without. We recommend a 30-seconds time limit for Q2, Q4, and Q6 and a 60-seconds time limit for Q3. In supplementary material, we provide details of how we chose these time limits. 
Figure~\ref{fig:recommendHist} shows the distribution over the 6 recommended questions of correct solutions in the groups with the time limits applied.

\section{Conclusion}
In previous online studies with programmers, researchers often relied on participants' claims to have programming skill or they used programming tasks or  programming knowledge questions to verify these. 
Our work showed, however, that designing programming screener questions is not trivial and we would not recommend using questions without testing them before. While we raised a methodological problem in software-engineering work on a meta-level, we also contributed concrete and validated screener questions on a primary level.

We surveyed a total of 249 people to find questions that can be used to filter participants with programming skill. To get a ground truth sample of programmers, we selected participants for whom we were able to verify that they actually have any programming skill. We, then, recruited non-programmers and Clickworkers with and without self-reported programming skill to test our screening instrument. 
Finally, we tested our instrument under adversarial conditions to test its robustness. Based on our evaluation, we recommend 6 of our 16 screener questions for use in online studies. 

In future work, we will continue to expand and test our question set. While the small set is sufficient to protect against non-adversarial participants, who simply have a different interpretation what programming is than we do, a larger set will be more robust in an adversarial setting.



\section*{Acknowledgments}
This work was partially funded by the ERC Grant 678341: Frontiers of Usable Security. 
\bibliographystyle{IEEEtran}
\bibliography{bib}

\end{document}


\title{Supplementary Material: Do you \emph{really} code? - Designing and Evaluating Screening Questions for Online Surveys with Programmers}

\maketitle

\section{Structure}
\label{appendix:structure}
Our supplementary material is structured as follows.
\begin{enumerate}
    \item \textbf{Section~\ref{appendix:structure}} outlines the structure of the supplementary material.
    \item In \textbf{Section~\ref{appendix:survey}}, we provide information on our survey questions. 
    \item In \textbf{Section~\ref{appendix:Demographics}} we give details on the demographics of our participants.
    \item \textbf{Section~\ref{appendix:results}} summarizes results on the effectiveness and efficiency of our tested screener questions. Additionally, this section also shows an overview of information resources our participants from the attack scenario used to answer the programming questions. 
    \item \textbf{Section~\ref{appendix:TimingThresholds}} reports the timing thresholds for the 4 screener questions Q2, Q3, Q4, and Q6, which we recommended to use with time limits. 
\end{enumerate}

\section{Survey}
\label{appendix:survey}
The correct answers are marked with $\text{\rlap{$\checkmark$}}\square$. All questions were randomly shown to the participants. 
\includegraphics[width=0.5\textwidth]{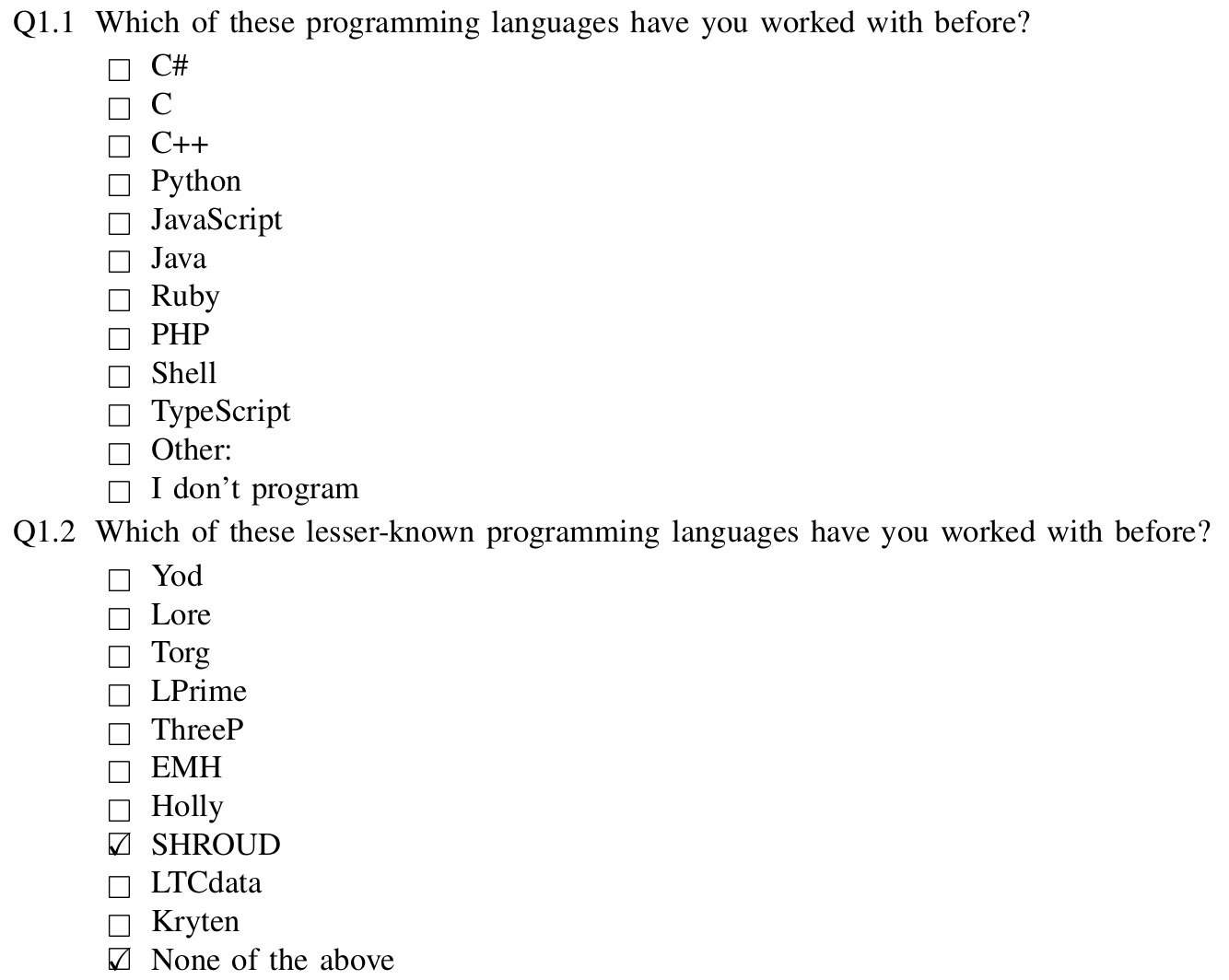}
\includegraphics[width=0.5\textwidth]{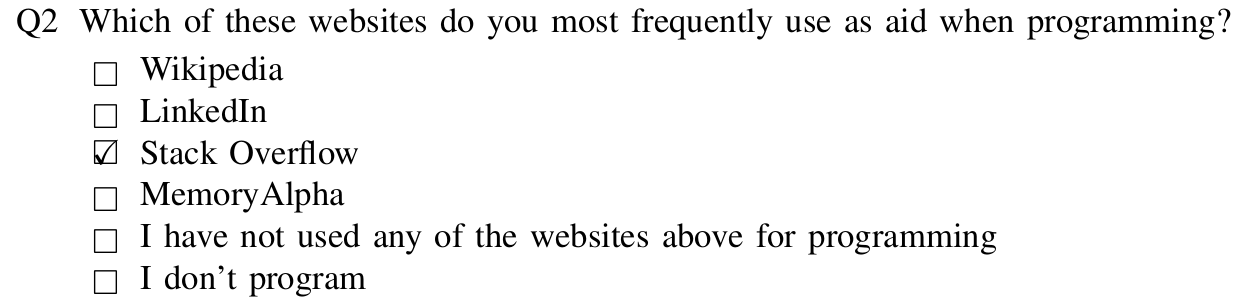}
\includegraphics[width=0.5\textwidth]{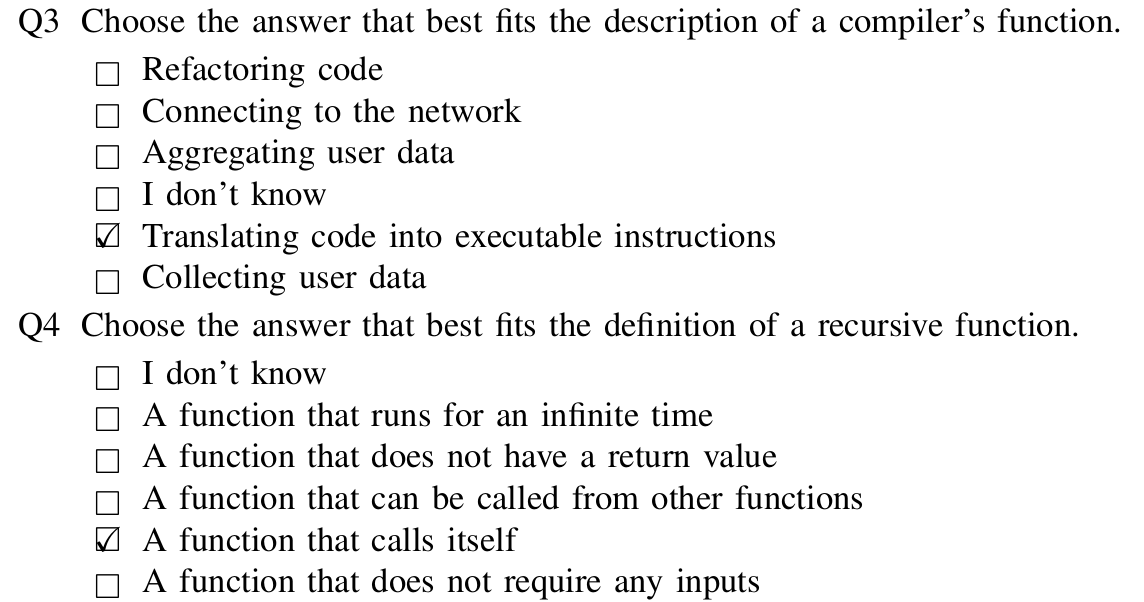}
\includegraphics[width=0.5\textwidth]{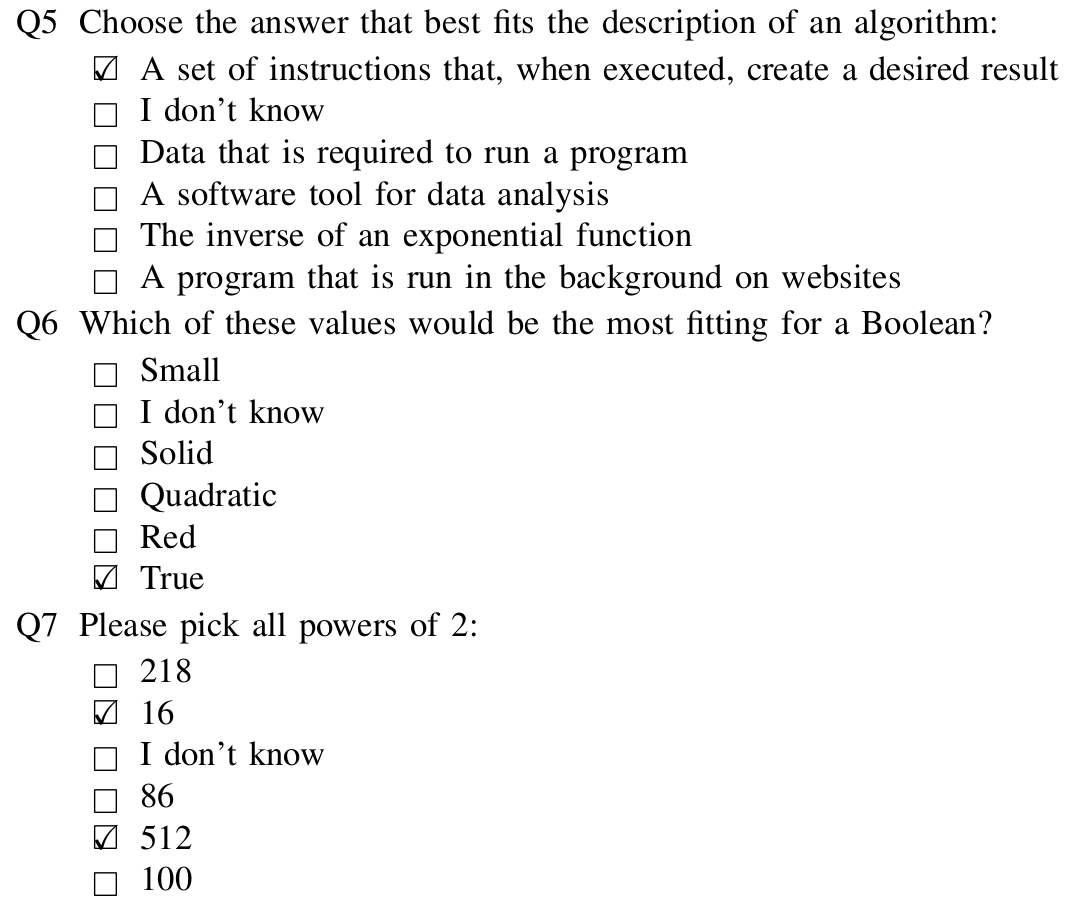}
\includegraphics[width=0.5\textwidth]{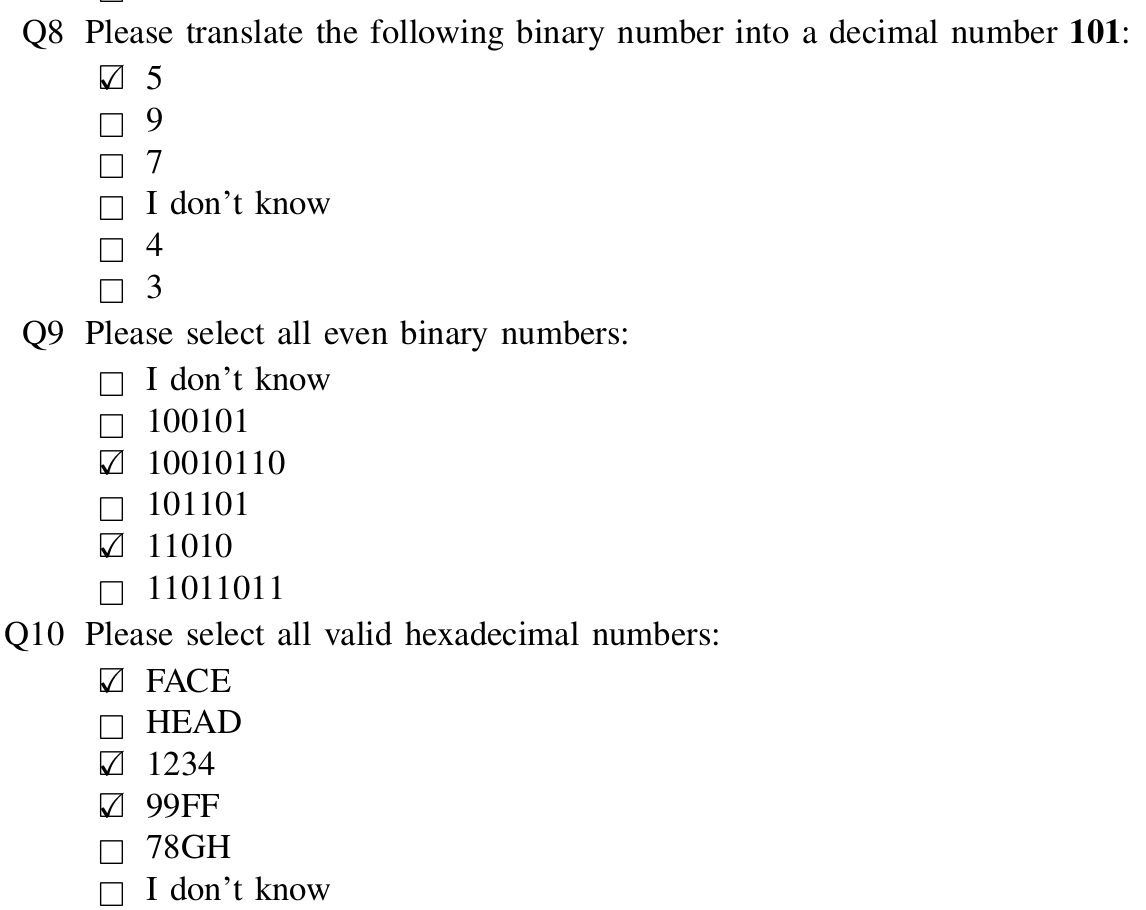}
\includegraphics[width=0.4\textwidth]{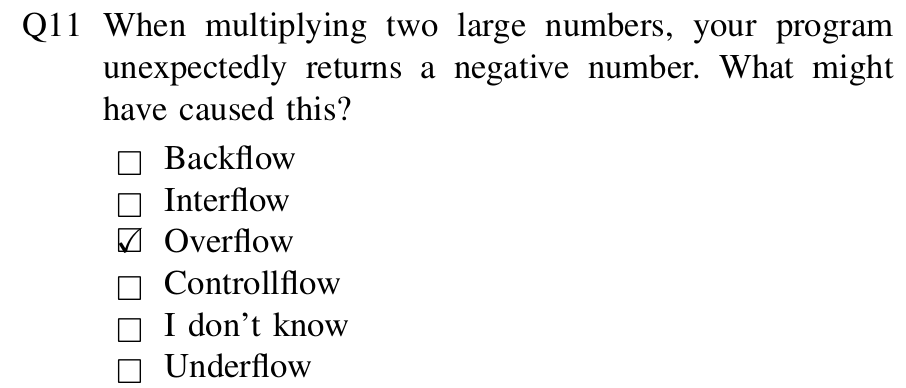}
\includegraphics[width=0.3\textwidth]{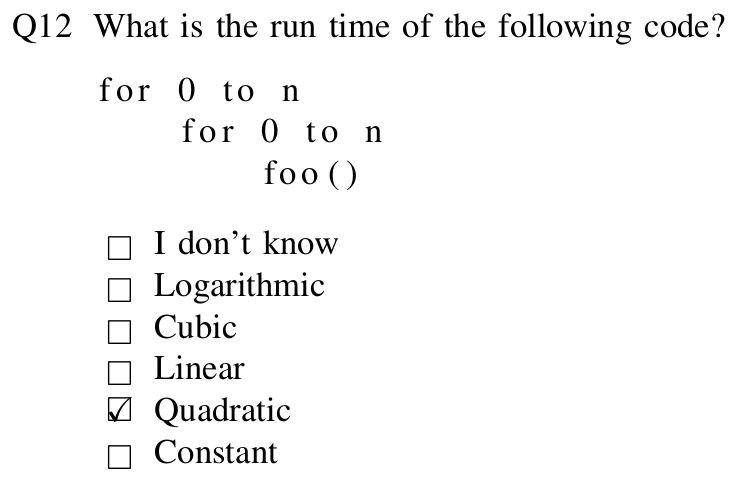}
\includegraphics[width=0.45\textwidth]{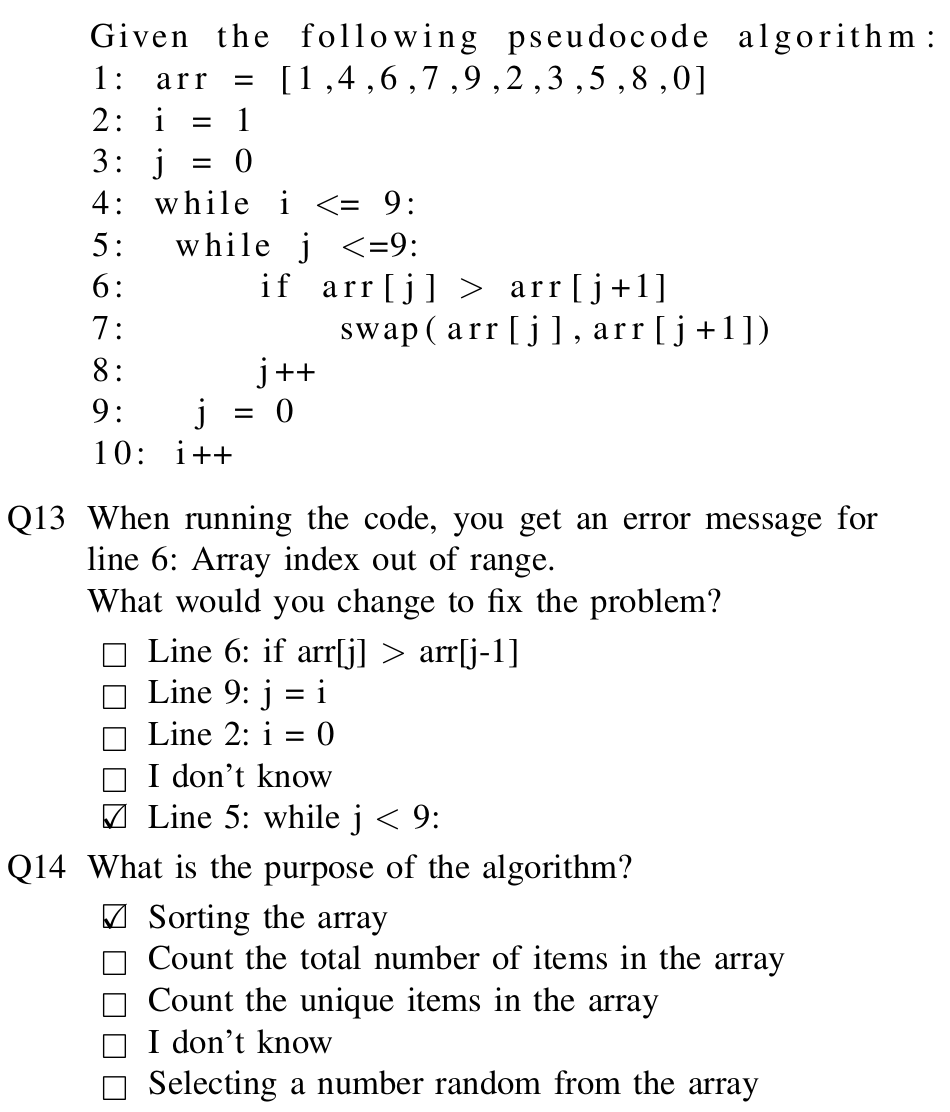}
\includegraphics[width=0.45\textwidth]{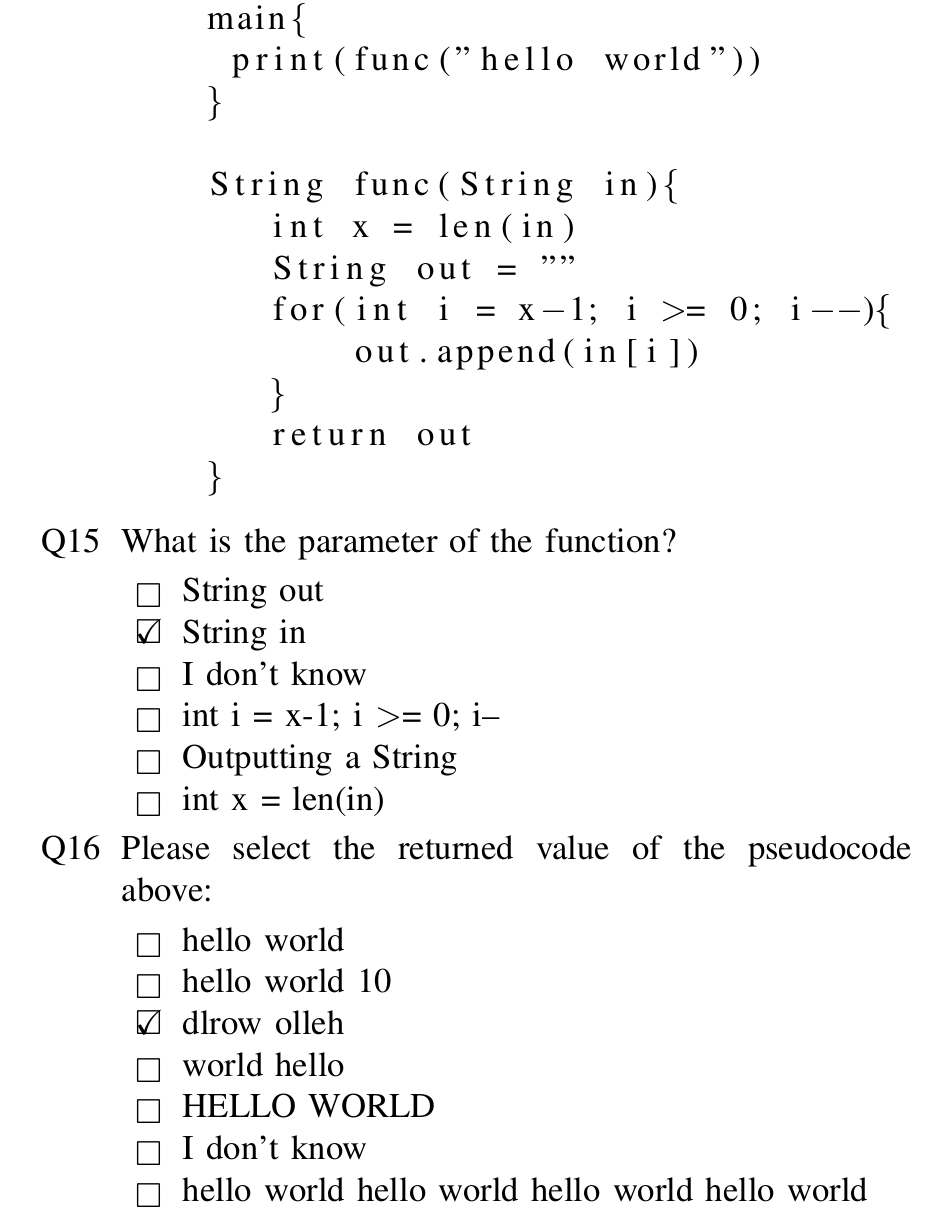}

\begin{enumerate}
    \item \textbf{[Attention Check]} This is an attention check question. Please select the
answer "Octal". \begin{itemize} \renewcommand{\labelitemi}{\raisebox{-.25\height}{$\square$}}
    \item Duodecimal
    \item I don't know
  \item [$\text{\rlap{$\checkmark$}}\square$] Octal
    \item Binary
    \item Decimal
    \item Hexadecimal
\end{itemize}{}

\end{enumerate}

\subsection*{Demographic Questions}
\begin{enumerate}
 
    \item How many years of programming experience do you
have?

\item How experienced would you consider yourself at
programming?
\begin{itemize} \renewcommand{\labelitemi}{\raisebox{-.25\height}{$\square$}} 
\item Beginner
\item Intermediate
\item Expert
\item No experience at all
\end{itemize}

\item Have you ever been paid for your work as a
programmer?
\begin{itemize} \renewcommand{\labelitemi}{\raisebox{-.25\height}{$\square$}} 
\item Yes
\item No
\end{itemize}

\item In which country do you currently reside?
\item How old are you?
\item What is your gender?
\begin{itemize} \renewcommand{\labelitemi}{\raisebox{-.25\height}{$\square$}}
\item Male
\item Female
\item Prefer to self-describe:
\item Prefer not to tell
\end{itemize}
\item What is your main profession?

\end{enumerate}{}

\section{Participants' Demographics}
\label{appendix:Demographics}
\subsection{Country of Residence}

\subsubsection*{\textbf{Clickworker No Experience:}}
NA:           (1), Finland (1),
 France (1),
 Germany (21),
 India (3),
 Italy (1),
 Netherlands (1),
 Philippines (1),
 Russian Federation (1),
Serbia (1),
South Africa (2),
Spain (2),
United Kingdom of Great Britain and Northern Ireland (8),
United States of America (6)

\subsubsection*{\textbf{Clickworker with Programming Experience:}}
 Argentina (1),
 Austria (1),
 Egypt (2),
 Finland (1),
 Germany (22),
 Greece (1),
 India (4),
 Indonesia (1),
 Italy (4),
Kenya (1),
Netherlands (1),
 Nigeria (1),
Portugal (1),
Romania (1)
Spain (1),
Sweden (1),
United Kingdom of Great Britain and Northern Ireland (5),
United States of America (3)

\subsubsection*{\textbf{Clickworker Attack:}}
Australia (1), Austria (1), Brazil (3), Colombia (1), Finland (1), France (1), Germany (16), India (1), Italy (4), Kenya (2), Malaysia (1), Mexico (1), Peru (1), Poland (2), Romania (1), Spain (4), Sweden (1), United Kingdom of Great Britain and Northern Ireland (1), United States of America (4)

\subsection{Main Occupation}
\subsubsection*{\textbf{Professional Developers:}}
Lead Developer (3), Software Developer (22), IT Staff (2), Data scientist (1), Engineer (1), IT Release Manager (1), System architect (1), Function Developer (1), Security Consultant (1)

\subsubsection*{\textbf{Clickworker without Programming Skill:}}
Student (4),
House wife (3), Freelancer (3),  Engineer (3),
Merchant (2), Teacher (2),Administrator (2), Developer (2), Editor (2), 
Social worker (1), 
Self-employed (1), 
Manager (1), 
Druggist (1), 
Scientist (1),
Project manager (1), 
Telephone Operator (1), 
Security (1), 
Office clerk (1), 
Public administration specialist (1), 
Bookkeeping (1),
Paralegal (1), 
Investment analysis (1), 
Doctor (1), 
Unemployed (1), 
Physicist (1),  
Translator (1), 
Geologist (1), 
Tattoo artist (1), 
IT Security (1), 
Economist (1), 
Environmental Health Officer (1), 
Controller  (1), 
Service staff (1), 
Support (1), 
Finance (1),
NA (1)

\subsubsection*{\textbf{Clickworker with Programming Skill:}}
Student (9),
Software Developer (8), 
Support (7),
IT (4),
Self-employed (2), 
House wife/husband (2),
Law (1),   Sales (1),  Spiritual life coaching (1), Nurse (1), Data scientist (1), Office employee (1), Engineer (1), System administrator (2), Scientist (4), Chemist (1), Assistant lecturer (1), Unemployed (1),  Graphic designer (1), Architecture (1), Astronomer (1), Nutritionist (1)

\subsubsection*{\textbf{Clickworker Attack:}}
Student(4), Clerk (3), Sales (3), Manager (2), Teacher (2), Worker (2), Administrative (1), Architecture (1), Asacom (1),
Business (1), Carrier (1), Civil engineering (1), Crowdworker (1), 
Data handling (1), Data scientist (1), Delivery driver (1), 
Developer (1), Factory technician (1), Finance (1), Freelancer (1), 
IT (1), Landscape architecture (1), Editor (1),
Material technician (1), Mechanic (1), Medical doctor (1), Media (1), Musician (1), Online shop operator (1), 
Physiotherapist (1), Programming (1), Independent (1), Software developer (1), 
Social pedagogue (1), Student developer (1), Student assistant (1), NA (1)



\section{Results}
\label{appendix:results}

\subsection{Effectiveness}
\begin{figure}[]
\centering
   \includegraphics[width=0.5\textwidth]{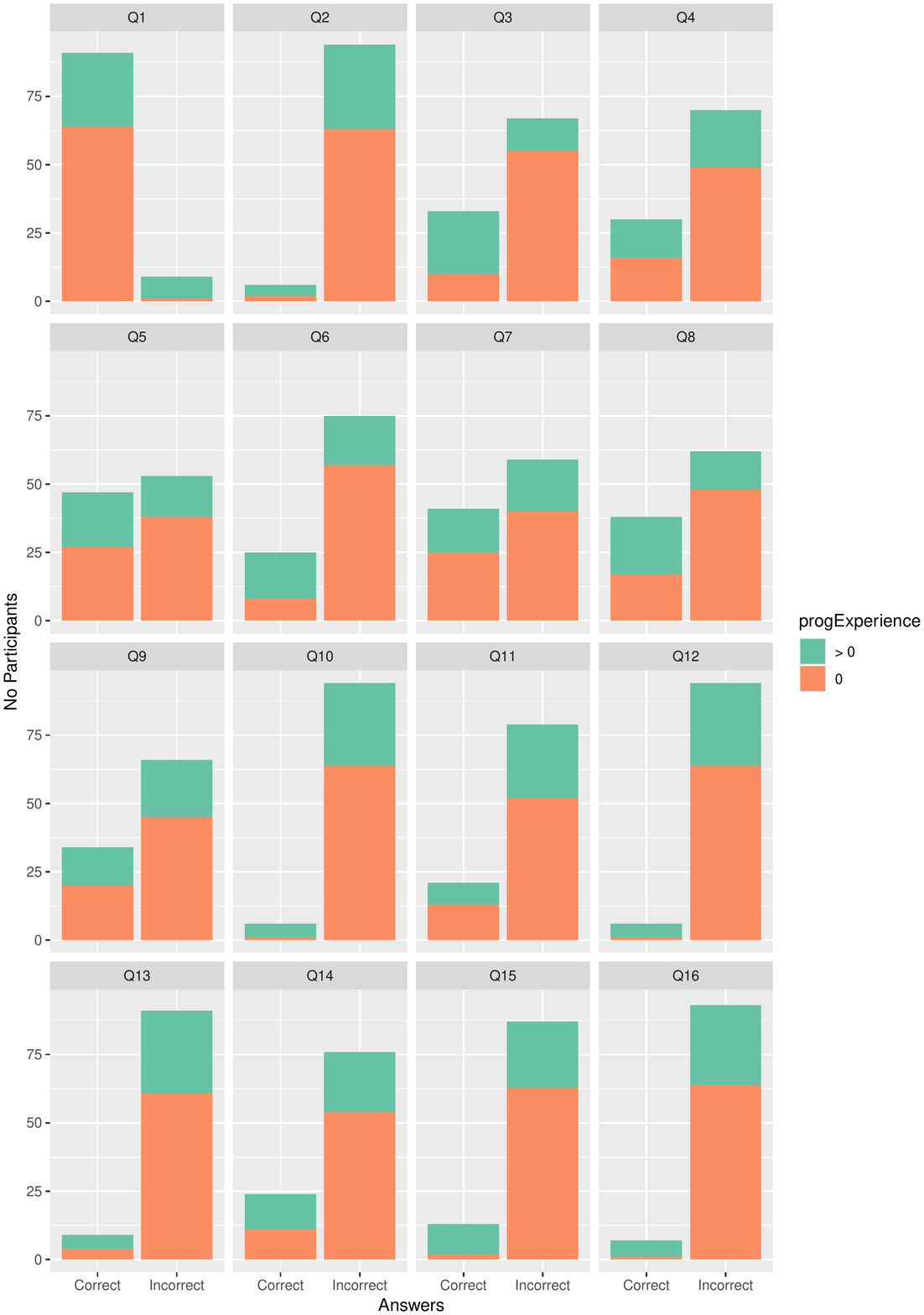}
        \caption{Number of correct solutions of the non-programmer group (n = 100) separated by participants who indicated to have 0 years and more than 0 years of programming experience.}
       \small
       \textbf{progExperience}: $>$ 0 if non-programmer participants reported to have more than 0 years of programming experience, and = 0 if they reported to have no programming experience at all.
       \label{fig:expQuestion}
\end{figure}

Table~\ref{table:allquestions} shows an overview of the correct and incorrect answers of the programmer and non-programmer group for all the questions (Q1-Q16). Table~\ref{table:statisticsAllQuestions} summarizes the related statistical analysis results. Figure~\ref{fig:expQuestion} displays the number of correct solutions of the non-programmer group (n = 100) separated by participants who indicated to have 0 years and more than 0 years of programming experience.

\subsection{Efficiency}
\begin{figure}[]
\centering
    \includegraphics[width=0.5\textwidth]{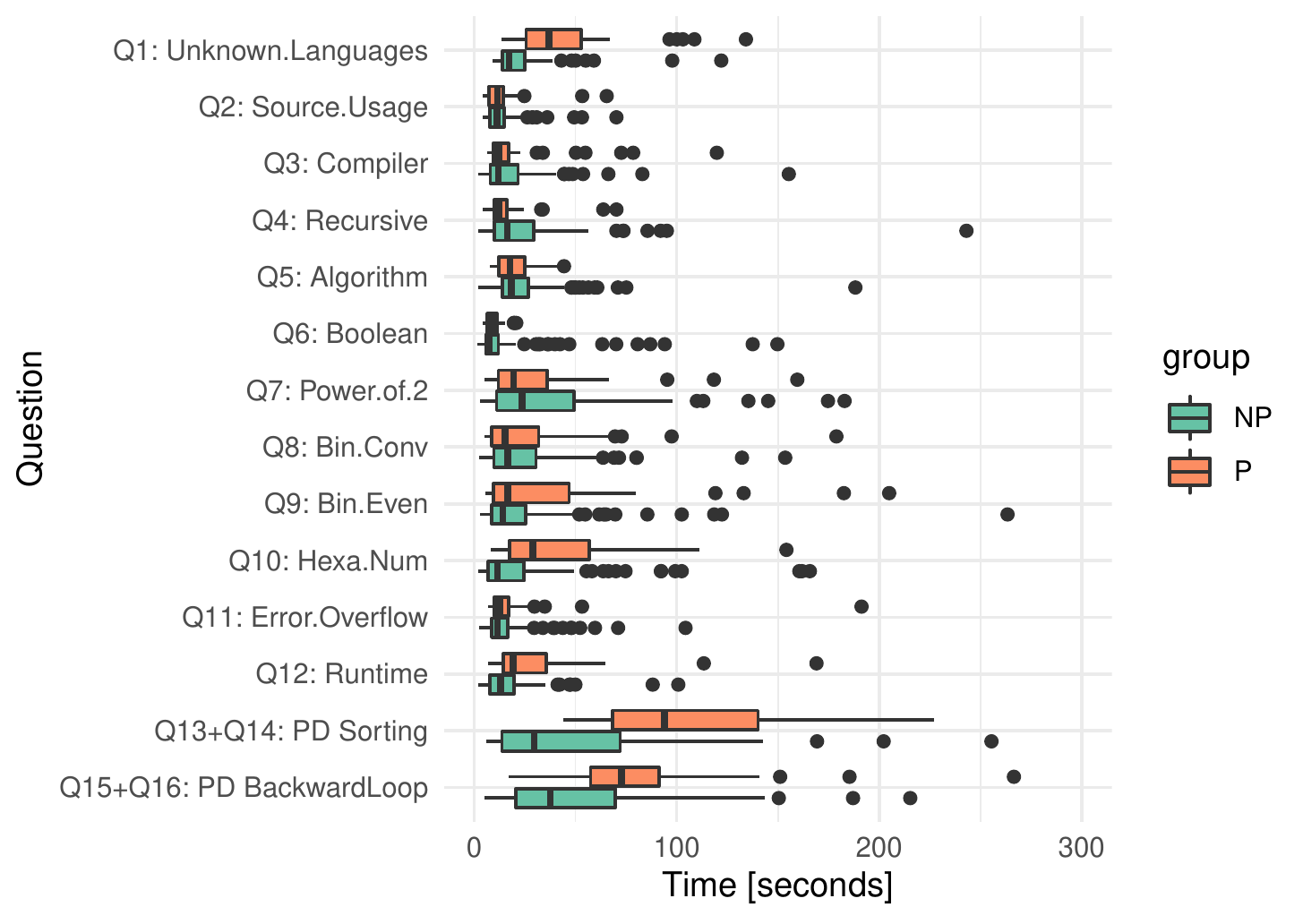}
        \caption{Time to answer each question for the non-programmer (NP) and programmer (P) group.}
        \label{fig:time}
\end{figure}

We visualized the mean times for both programmer and non-programmer groups according to each task block in Figure~\ref{fig:time}.

\subsection{Attack Scenario}
Table~\ref{table:resources} shows an overview of the self-reported resources used by participants for correctly answering the questions within the attack scenario (n = 47).

\section{Timing thresholds}
\label{appendix:TimingThresholds}
\begin{figure}[h]
\centering
   \includegraphics[width=0.45\textwidth]{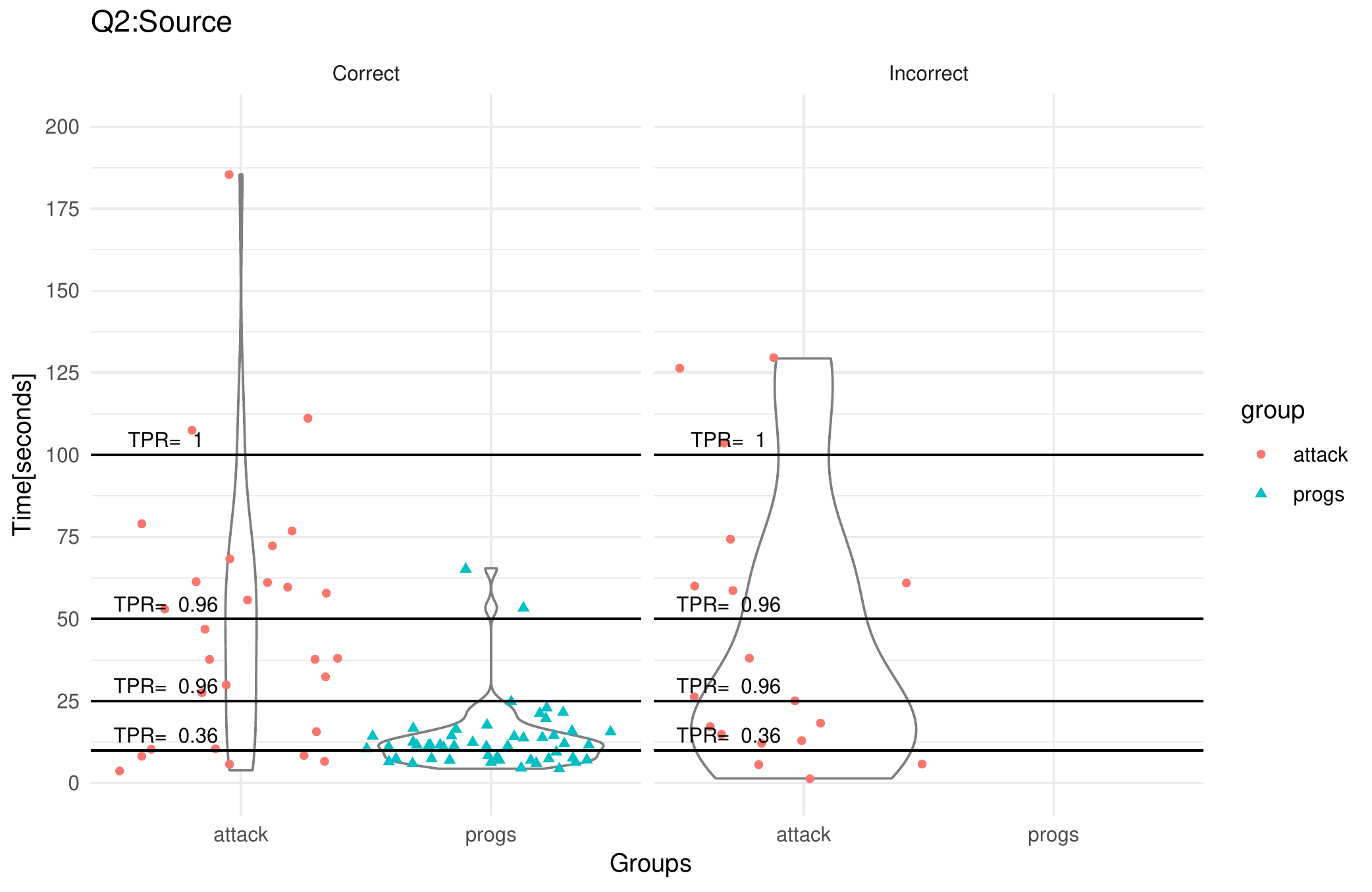}
        \caption{}
       \small
       The figure illustrates the true positive rate at example thresholds (10 seconds, 25 seconds, 50 seconds, 100 seconds)
       \label{fig:t1}
\end{figure}

\begin{figure}[h]
\centering
   \includegraphics[width=0.4\textwidth]{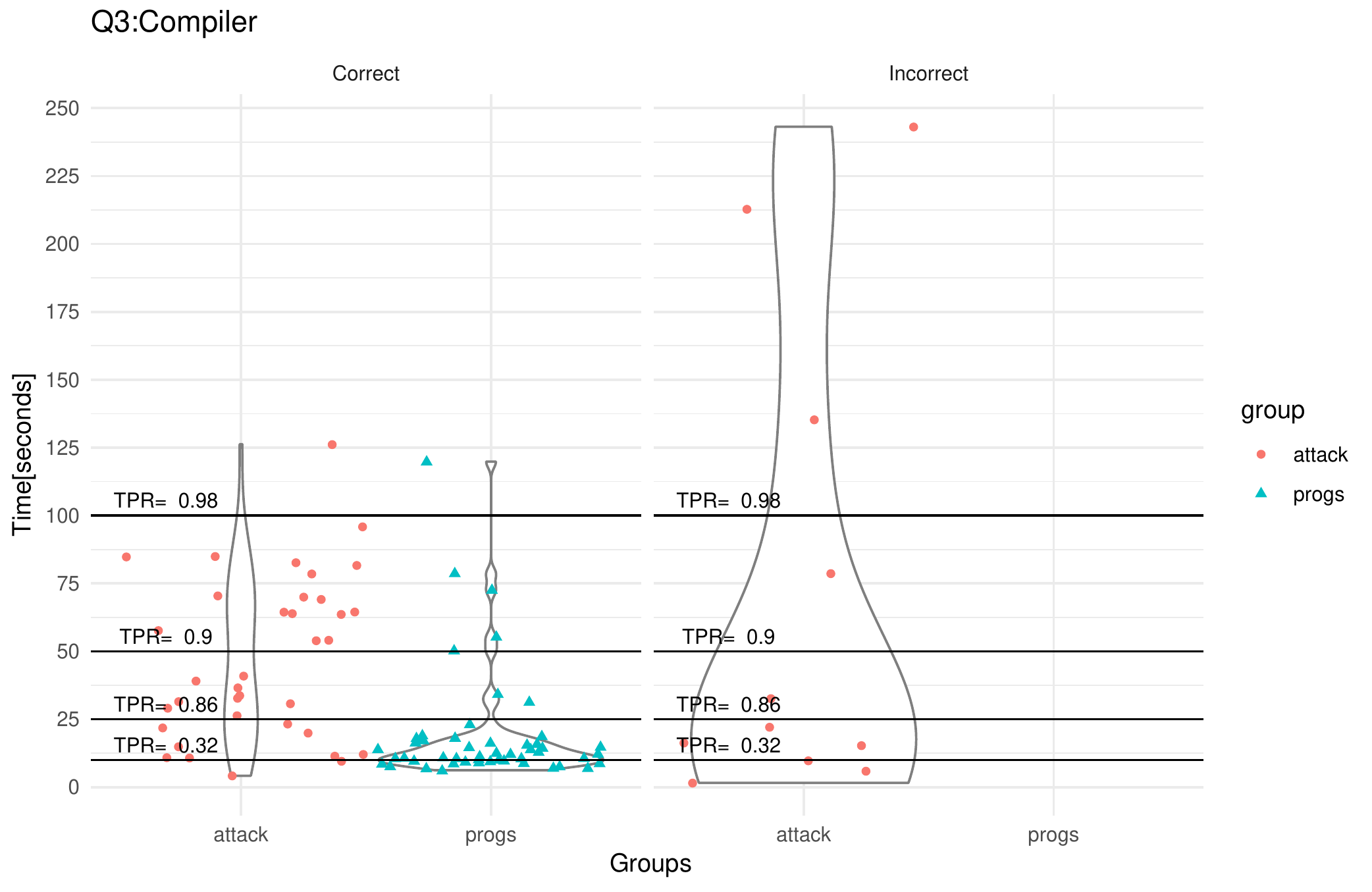}
        \caption{}
       \small
       The figure illustrates the true positive rate at example thresholds (10 seconds, 25 seconds, 50 seconds, 100 seconds).
       \label{fig:t2}
\end{figure}

\begin{figure}[h]
\centering
   \includegraphics[width=0.4\textwidth]{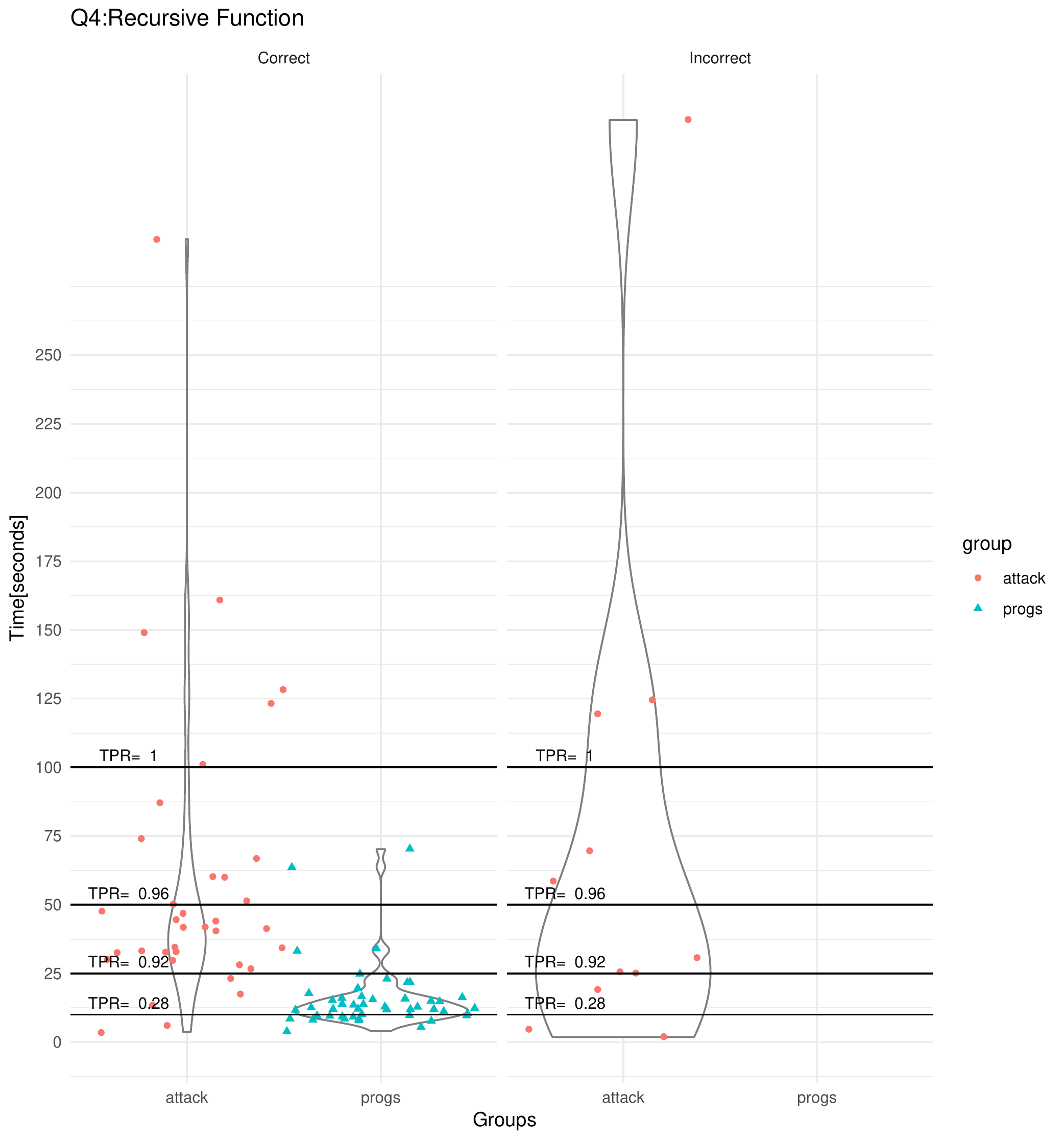}
        \caption{}
       \small
       The figure illustrates the true positive rate at example thresholds (10 seconds, 25 seconds, 50 seconds, 100 seconds)
       \label{fig:t3}
\end{figure}

\begin{figure}[h]
\centering
   \includegraphics[width=0.4\textwidth]{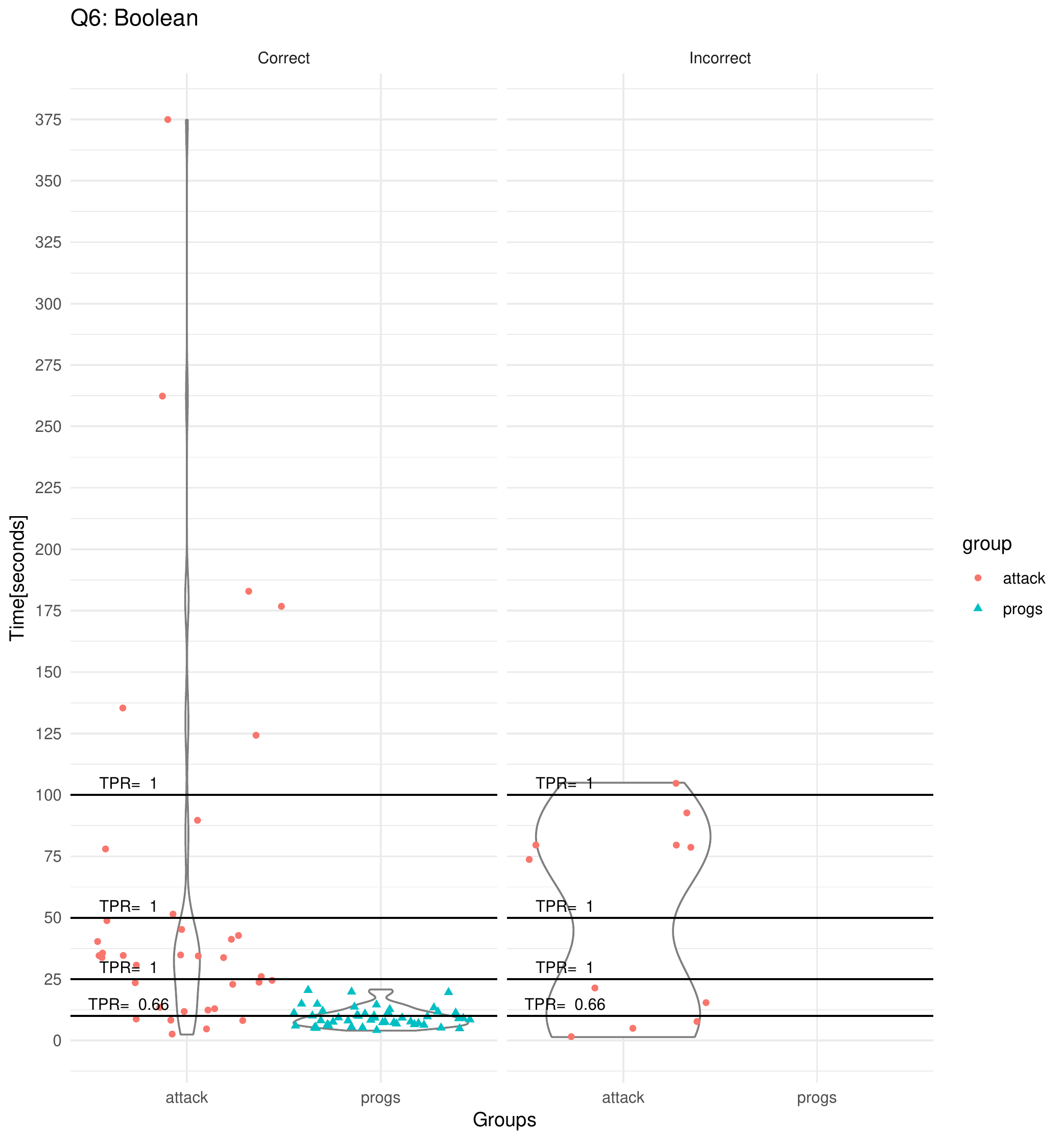}
        \caption{}
       \small
       The figure illustrates the true positive rate at example thresholds(10 seconds, 25 seconds, 50 seconds, 100 seconds)
       \label{fig:t4}
\end{figure}

Figures~\ref{fig:t1}, \ref{fig:t2}, \ref{fig:t3}, and~\ref{fig:t4}, illustrate how many participants from the attacker (attack) and programmer group (progs) solved the four recommended questions Q2, Q3, Q4, and Q6 correctly and how many seconds they needed to answer.
Drawing a line on a certain time threshold, it becomes visible how many attackers and programmers would be excluded from the correct group.

\begin{table}[]
   \begin{subtable}{.5\linewidth}
    \begin{tabular}{c|cc}
    & \cmark& \xmark\\
        \midrule
         Programmer&49&1 \\
         Non-Programmer&91&9\\ 
    \end{tabular} 
    \caption{Unknown.Languages (Q1)}
    \label{subtable:Unknown.Languages}
     \end{subtable}
\begin{subtable}{.38\linewidth}   
    \begin{tabular}{c|cc}
    & \cmark& \xmark~(Idp)\\
        \midrule
         Programmer&50 &0 (0) \\
         Non-Programmer&6&94 (60)\\
    \end{tabular}    
    \caption{Source.Usage (Q2)}
    \label{subtable:Source.Usage}
    \end{subtable}
    \newline\\\\
   \begin{subtable}{.5\linewidth}
    \begin{tabular}{c|cc}
    & \cmark& \xmark~(Idk)\\
        \midrule
         Programmer&50 &0 (0)\\
         Non-Programmer&33&67 (37)\\ 
    \end{tabular}    
    \caption{Compiler (Q3)}
    \label{subtable:Compiler}
     \end{subtable}
\begin{subtable}{.45\linewidth}   
    \begin{tabular}{c|cc}
    & \cmark& \xmark~(Idk)\\
        \midrule
         Programmer&50 &0 (0) \\
         Non-Programmer&30&70 (33)\\
    \end{tabular}    
    \caption{Recursive (Q4)}
    \label{subtable:Recursive}
    \end{subtable}
    \newline\\\\
\begin{subtable}{.5\linewidth}   
    \begin{tabular}{c|cc}
    & \cmark& \xmark~(Idk)\\
        \midrule
         Programmer&49  &1 (0)\\
         Non-Programmer&47&53 (13)\\
    \end{tabular}    
    \caption{Algorithm (Q5)}
    \label{subtable:Algorithm}
    \end{subtable}    
\begin{subtable}{.4\linewidth}   
    \begin{tabular}{c|cc}
    & \cmark& \xmark~(Idk)\\
        \midrule
         Programmer&50  &0 (0)\\
         Non-Programmer&25&75 (60)\\
    \end{tabular}    
    \caption{Boolean (Q6)}
    \label{subtable:Boolean}
\end{subtable}  
\newline\\\\
   \begin{subtable}{.5\linewidth}
    \begin{tabular}{c|cc}
    & \cmark& \xmark~(Idk)\\
        \midrule
         Programmer&48 &2 (0) \\
         Non-Programmer&41&59 (18)\\ 
    \end{tabular}    
    \caption{Power.of.2 (Q7)}
    \label{subtable:Power.of.2}
     \end{subtable}
\begin{subtable}{.4\linewidth}   
    \begin{tabular}{c|cc}
    & \cmark& \xmark~(Idk)\\
        \midrule
         Programmer&48 &2 (0) \\
         Non-Programmer&38&62 (38)\\
    \end{tabular}    
    \caption{Bin.Conv (Q8) } 
    \label{subtable:Bin.Conv}
    \end{subtable}
\newline\\\\
\begin{subtable}{.5\linewidth}   
    \begin{tabular}{c|cc}
    & \cmark& \xmark~(Idk)\\
        \midrule
         Programmer&45  &5 (2)\\
         Non-Programmer&34&66 (39)\\
    \end{tabular}    \caption{Bin.Even (Q9)} \label{subtable:bin.even}
    \end{subtable}    
\begin{subtable}{.4\linewidth}   
    \begin{tabular}{c|cc}
    & \cmark& \xmark~(Idk)\\
        \midrule
         Programmer&35  &15 (0)\\
         Non-Programmer&6&94 (51)\\
    \end{tabular} \caption{Hexa.Num (Q10)} \label{subtable:Hexa.Num}
\end{subtable}  
\newline\\\\
   \begin{subtable}{.5\linewidth}
    \begin{tabular}{c|cc}
    & \cmark& \xmark~(Idk)\\
        \midrule
         Programmer&47&3 (2) \\
         Non-Programmer&21&79 (49)\\ 
    \end{tabular}    
    \caption{Error.Overflow (Q11)} 
    \label{subtable:error.overflow}
     \end{subtable}
\begin{subtable}{.4\linewidth}   
    \begin{tabular}{c|cc}
    & \cmark& \xmark~(Idk)\\
        \midrule
         Programmer&40 &10 (5)\\
         Non-Programmer&6&94 (56)\\
    \end{tabular}    
    \caption{Runtime (Q12)} 
    \label{subtable:runtime}
    \end{subtable}
\newline\\\\
   \begin{subtable}{.5\linewidth}
    \begin{tabular}{c|cc}
    & \cmark& \xmark~(Idk)\\
        \midrule
         Programmer&38&12 (2) \\
         Non-Programmer&9&91 (66)\\ 
    \end{tabular}    
    \caption{Error.OutOfBound (Q13)} 
    \label{subtable:error.outofbound}
     \end{subtable}
    \begin{subtable}{.49\linewidth}   
    \begin{tabular}{c|cc}
    & \cmark& \xmark~(Idk)\\
        \midrule
         Programmer&49 &1 (0)\\
         Non-Programmer&24&76 (51)\\
    \end{tabular}    
    \caption{Sorting.Array (Q14)} 
    \label{subtable:sortingarray}
    \end{subtable}
    \newline\\\\
   \begin{subtable}{.5\linewidth}
    \begin{tabular}{c|cc}
    & \cmark& \xmark~(Idk)\\
        \midrule
         Programmer&50&0 (0) \\
         Non-Programmer&13&87 (57)\\
    \end{tabular}    
    \caption{Function.Param (Q15)}
    \label{subtable:Function.Param}
     \end{subtable}
    \begin{subtable}{.4\linewidth}   
    \begin{tabular}{c|cc}
    & \cmark& \xmark~(Idk)\\
        \midrule
         Programmer&47 &3 (0) \\
         Non-Programmer&7&93 (41)\\
    \end{tabular}    
    \caption{Backward.Loop (Q16)} 
    \label{subtable:backward.loop}
    \end{subtable}
    \caption{Number of (in)correct answers of the programmers~(n = 50) and non-programmers (n = 100) for Q1-Q16.}
    \centering
    Idp = I don’t program; Idk = I don't know.
    \label{table:allquestions}
\end{table}

\begin{table*} []\centering \footnotesize \begin{adjustbox}{max width=\textwidth, max totalheight=\textheight}
\begin{tabular}{cccc}
\toprule
\textbf{Question}&\textbf{O.R.}&\textbf{CI}&\textbf{p-value}\\ 
\midrule
Q1&4.8&[0.63, 216.47]&0.17\\ 
Q2&Inf&[143.54, Inf]&$<$  0.001$^{*}$\\ 
Q3&Inf&[23.93, Inf]&$<$  0.001$^{*}$\\ 
Q4&Inf&[27.37, Inf]&$<$ 0.001$^{*}$\\ 
Q5&54.18&[8.56, 2240.58]&$<$ 0.001$^{*}$\\ 
Q6&Inf&[34.71, Inf]&$<$ 0.001$^{*}$\\ 
Q7&33.82&[8.09, 302.74]&$<$ 0.001$^{*}$\\ 
Q8&38.29&[9.14, 343.35]&$<$ 0.001$^{*}$\\ 
Q9&17.11&[6.05, 60.28]&$<$ 0.001$^{*}$\\ 
Q10&35.14&[12.11, 120.60]&$<$ 0.001$^{*}$\\ 
Q11&56.86&[15.96, 310.37]&$<$ 0.001$^{*}$\\ 
Q12&59.35&[19.37, 218.09]&$<$ 0.001$^{*}$\\ 
Q13&30.79&[11.45, 92.27]&$<$ 0.001$^{*}$\\ 
Q14&149.39&[23.1, 6075.50]&$<$ 0.001$^{*}$\\ 
Q15&Inf&[71.76, Inf]&$<$ 0.001$^{*}$\\ 
Q16&190.22&[45.97, 1170.64]&$<$ 0.001$^{*}$\\ 
\bottomrule
\end{tabular}
\end{adjustbox} 
\caption{Summary of statistical analysis for questions Q1-Q16 for the programmer and non-programmer group.}
\small{Fisher's exact tests were used for analysis. The independent variable was the programmer/non-programmer group. The dependent variable was the correctness of an answer. Significant results are marked with $^{*}$.} 
\label{table:statisticsAllQuestions} 
\end{table*} 

\begin{table*} []\centering \footnotesize \begin{adjustbox}{max width=\textwidth, max totalheight=\textheight} \begin{tabular}{cc|ccl}
\toprule
\textbf{Question}&\textbf{No of correct responses} &\textbf{Used Internet search} &\textbf{Asked friends/colleagues}& \textbf{Other} \\
\midrule
Q2&29&12&4&0\\
Q3&36&14&4&1: "I actually knew that myself! :)"\\
Q4&36&19&3&0\\
Q6&36&16&2&1:"I knew it could have a value of true or false"\\
Q14&29&0&1&1:"Guessed"\\
Q15&13&0&0&0\\
Q16&12&1&2&0\\

\bottomrule
\end{tabular} 
\end{adjustbox} 
\caption{Self-reported resources used by participants for correctly answering the questions within the attack scenario (n = 47). Multiple answers were possible.} 
\label{table:resources} 
\end{table*}



